\documentclass[twocolumn, prb, superscriptaddress,
  showpacs]{revtex4-1} 
\usepackage{amsmath, array} 
\usepackage{graphicx}
\usepackage{grffile}
\usepackage{hyperref} 
\usepackage{color}
\usepackage{xcolor}

\newcommand{\EqLabel}[1]{\label{#1}} 

\begin{document}
\title{Particularities of polaron formation in the extended Holstein model with next nearest neighbor transfer}
\author{Monodeep Chakraborty} \email{bandemataram@gmail.com} \affiliation{Centre for Quantum Science and Technology, Chennai Institute of Technology, Chennai-600069, India}
\author{Jayita Chatterjee} \email{jayita.chatterjee@sncwgs.ac.in} \affiliation{Department of Physics, Sarojini Naidu College for Women, 30 Jessore Road,
Kolkata–700028, India}
\author{Holger Fehske}\email{fehske@physik.uni-greifswald.de}
	\affiliation{Institute of Physics, University of Greifswald, Felix-Hausdorff-Straße 6, 17489 Greifswald, Germany}
	\affiliation{Erlangen National High Performance Computing Center, 91058 Erlangen, Germany}
\begin{abstract}
Employing a largely unbiased variational exact diagonalization technique, we analyze the consequences of longer-ranged electron hopping and electron-phonon interaction on polaron formation in one dimension. Having at our disposal the accurate  ground-state energy and wave function, we calculate and discuss 
various physical quantities, such as the renormalized band structure, effective mass, wave-function renormalization factor, phonon dressing, and Drude weight, 
characterizing the properties of the polaronic quasiparticle. We demonstrate that the electron-phonon coupling affects the relative strength  of the 
nearest neighbor (NN) and next nearest neighbor (NNN) hopping processes
in a dynamic way. 
Most notably we observe that the minimum of the polaron band, occurring at a finite 
momentum for large negative ratio between NN and NNN transfer, jumps to zero momentum as the electron-phonon coupling exceeds a critical one, thereby causing a rather sharp polaron transition in the one-dimensional extended  Holstein model. The signatures of this transition are seen in the effective mass and  polaron mobility, and therefore should be easily detectable by transport measurements.  
\end{abstract}
\date{\today}
\maketitle

\section{Introduction}
Electron-phonon (EP) coupling causes many fascinating phenomena in solid state systems. In particular, a strong mutual interaction between the charge carriers and lattice deformations can ensure that new quasiparticles, electrons dressed by phonon clouds, form. These entities, referred to as polarons, are characterized by a substantially increased effective mass, as well as spectral, optical, and transport properties that notably differ from those of normal band carriers \cite{LA33,MG40,Fi75,SS93,Fi95,FT07}. The microscopic structure of polarons is very diverse. Here the form of the particle-phonon coupling, the lattice and  band structures, and the spatial dimension  come into play. Depending on the system and thus physical conditions polarons can be small (Holstein type \cite{Ho59a,Ho59b}) or large (Fröhlich type \cite{Fr54,DE06}). Extended Holstein \cite{AK99,FLW00}, Edwards \cite{AEF07,CMTMF16}, Rashba-Pekar \cite{Ra82} or pseudo Jahn-Teller \cite{PW84,EBKT03} polarons are other species.  The polaron problem addressed as early as 1933 has received  renewed attention due to the observation of polaronic effects in new material classes exhibiting exceptional properties \cite{IRF06,Eg06,SAL95,Caea97,ALex07}. Examples are high-temperature superconducting cuprates \cite{BEMB92,Mott93}, charge ordered nickelates \cite{BE93,TSPD23}, or colossal magnetoresistive manganites \cite{MLS95,JHSRDE97,LE99}.

From a theoretical point of view the problem of ``self-trapping", which means the transition from rather mobile (weakly mass renormalized) polarons to quasi-immobile (heavily mass renormalized) polarons, has been discussed with much controversy over decades \cite{Ra82,GL91,WF98a}. We note that ``self-trapping" does not imply  a breaking of translational invariance, at least  at any finite phonon frequency. Since the lattice potential that tends to trap the carrier depends on the the carrier's state itself, self-trapping  is a complicated, highly nonlinear feedback phenomenon and most analytical approaches fail to describe the physically most interesting transition regime in the quantum phonon limit.  In recent years elaborate numerical approaches, such as  quantum and diagrammatic Monte Carlo \cite{RL82,BVL95,KP97,MPSS00,HEL04,BFM08,MNAFDCS09,Ho16}, exact diagonalization \cite{WRF96,St96,BTB99,KTB02,FT07,AFT08,AFT10,LBBM10,CMCD12,CTM14}, density matrix renormalization group \cite{ZJW98,JW98b,JF07} and kernel polynomial method \cite{wwaf06} based techniques, have been able to close this gap. According to their findings, the polaron transition is basically a {\sl continuous crossover} in the Holstein, extended Holstein, Edwards, and Fröhlich models; i.e., there no {\sl strict} self-trapping exists. The mass renormalization is especially large in systems, more specifically Holstein-type models, exhibiting a short-ranged  EP interaction with optical phonons. Thereby, in dimensions two and three, the crossover regime from light electrons to heavy polarons is relatively narrow. It appears at finite EP coupling strength and the heavy (Holstein) polarons formed are small.  The one-dimensional case is specific. Here large (heavy) polarons with a size of many lattice constants as well as small Holstein polarons may exist in the adiabatic regime of small phonon frequencies \cite{AFT10}. In the antiadiabatic limit of large phonon frequencies, polaron properties do not depend on dimension and the crossover is smoothened in general.

With regard to the central question of whether a sharp polaron transition may exist, a thorough analytical and numerical examination of the one-dimensional, single-particle Su-Schrieffer-Heeger (SSH) model \cite{SSH79} has brought to light that an EP coupling depending on both phonon and electron momenta can lead to a non-analyticity, as a function of the EP coupling parameter, in the polaronic properties \cite{MDCBNPMS10}. Most notably, at the critical coupling, the ground state changes suddenly from a state with zero momentum to one with finite momentum and the polaron mobility vanishes. We note that the main difference of the SSH model when compared to the standard Holstein- or Fröhlich -type models is that the phonons in the former directly cause bandwidth fluctuations and thereby generate longer-ranged electron hopping in a dynamical way \cite{MDCBNPMS10}.

 Against this background, we must ask ourselves whether the standard polaron lattice models with not only nearest neighbor (NN) hopping but direct longer-ranged electron transfer might also show a sharp polaron transition in a certain parameter regime; and whether such a transition will persist at finite phonon frequencies. For this purpose, in the present paper, we consider besides the Holstein model (HM) the extended Holstein model (EHM) with additional next nearest neighbor (NNN) transfer for one particle in one dimension. The importance of the NNN hopping in the HM was pointed out recently \cite{CDC11,CPM16}. The EHM takes also into account density-displacement-type long-range EP coupling to optical phonons and can be viewed as an extension of the Fröhlich model to a discrete lattice.  As compared to the HM, the EHM polaron is a large polaron in the whole EP coupling region. In addition, the polaron band is relatively weakly renormalized and the effective mass of the large EHM polaron is much smaller than that of the small HM polaron with the same polaron binding energy \cite{FLW00}.  Physically the EHM can be used, e.g., in order to model the interaction of doped holes with apical oxygens in the $\rm YBa_2Cu_3O_{6+x}$ high-$T_c$'s where the EP is not very screened \cite{AK99}.    These findings have renewed our interest in the 
one-dimensional EHM \cite{WCSMSD21,TMPSD21}.

The paper is organized as follows. In Sec.~II we introduce the models under consideration  and briefly review the exact numerical approach for their solution. In Sec.~III we present and discuss the numerical results. Section~IV contains a short summary and our conclusions.   

\section{Theoretical Model and Numerical Approach}
With our focus on polaron formation in  systems with a non-polar density-displacement-type (long-range) EP interaction, we consider the (extended) Holstein molecular crystal model
\begin{equation}
\EqLabel{H}
{\cal H}={\cal T} + {\cal H}_{\rm ph} + {\cal H}_{\rm e-ph}\
\end{equation}
for a one-dimensional lattice with lattice constant $a=1$. 

The first purely electron part of ${\cal H}$,
\begin{equation}
\EqLabel{T}
{\cal T} =-t_{1} \sum_{i}^{}\left( c_i^\dagger c_{i+1}^{} + {\rm H.c.}\right) \\
          -t_{2} \sum_{i}^{}\left( c_i^\dagger c_{i+2}^{} + {\rm H.c.}\right) \,
\end{equation}
describes the direct hopping of electrons between NN and NNN sites with amplitude $t_{1}$  and  $t_{2}$, respectively. The relevance of the NNN hopping in the HM has been demonstrated previously \cite{CDC11,CPM16}. 

The second purely phononic part, 
\begin{equation}
\EqLabel{P}
{\cal H}_{\rm ph} = \omega \sum_{i}^{} b_{i}^\dagger b_i^{}
\end{equation}
models a dispersionless (optical) Einstein mode of frequency $\omega$ (we set $\hbar=1$ throughout the paper).
In Eq.~\eqref{T}  [Eq.~\eqref{P}],  the fermionic [bosonic] operators $c_i^\dagger$ [$b_i^\dagger$] create an electron [phonon] at lattice site $i$, and $c_i^{}$ [$b_i^{}$] are the corresponding annihilation operators.

Finally, the EP-coupling is assumed to be cite{AK99,FLW00}
\begin{equation}
\EqLabel{EP}
 {\cal H}_{\rm e-ph}=-g\omega \sum_{ij}f_{j}(i) c_i^\dagger c_i^{} (b_{i+j}^{\dag}
+ b_{i+j}^{})
\end{equation}
with
\begin{equation}
f_{j}(i) = \frac{1}{[(i-j)^{2} +1 ]^{\frac{3}{2}}}\,,
\end{equation}
where the dimensionless EP coupling constant $g$ is related to the polaron binding energy: 
\begin{equation}
{\epsilon}_{p} =  g^2 \omega \sum_{j}f_{j}^{2}\,.
\end{equation} 
Recall that for the pure Holstein model, where $f_j(i)\equiv 1$, we have $g^2=\varepsilon_p/\omega$. 
$\sum_{j}f_{j}^{2} = 1.25$ for the EHM-3, where the EP interaction is spread over three sites, the site of the electron and its two NN sites. For the EHM-5, on the other hand, we have $\sum_{j}f_{j}^{2} = 1.266$; here the EP interaction spans over five sites. In the EHM-7 the range of the EP coupling comprises seven sites and  $\sum_{j}f_{j}^{2}= 1.268$. Adding up  $\sum_{j}f_{j}^{2}$ for the infinite lattice an approximate value of 1.27 will be obtained. Thus the density-displacement type long-range EP coupling  included in the EHM-7 fairly accounts for the  polaronic energy of a fully extended Holstein model.

Measuring in what follows the energy in units of $t_1$, the physics of the EHM with NNN transfer is governed by four ratios. The first ratio $\eta=t_2/t_1$ characterizes the bare band structure and, in particular, determines the location of the minimum of the band dispersion. Note that $\eta$ can be negative. The second so-called adiabaticity parameter 
$\alpha=2D\omega/W$ specifies which of the two subsystems, electrons or phonons, is the fast or  slow one, respectively. Accordingly $\alpha\ll1$ ($\alpha\gg 1$) classifies the adiabatic (anti-adiabatic) regime. Here $D$ is the spatial dimension and $W$ is the half-width of the bare electron band. From this we can deduce a third parameter $\lambda=\varepsilon_p/W$ which represents the ratio between polaron ``localization'' ($\propto\varepsilon_p$) and electron ``itinerancy" ($\propto W$). In the adiabatic regime, $\lambda$ serves as a parameter for the EP coupling strength; here the strong (weak) EP interaction regime is realized for  $\lambda\gg1$ ($\lambda\ll1$).  The fourth parameter $g^2$, which in a sense determines the relative deformation of the lattice that surrounds the particle (phonons dressing), can also be taken as a measure for the EP coupling strength. Obviously, $g^2$ is the relevant EP coupling parameter in the non-to-antiadiabatic regime, i.e., $g^2\gg1$ corresponds to the strong-coupling case. We will work with $g^2$ mainly because $W$ and therefore $\lambda$ noticeably depends  on $\eta$ in the $t_1$-$t_2$ EHM. 

The numerical treatment of the EHM is performed with optimized exact diagonalization (ED)  techniques based on the Lanczos algorithm \cite{CW85}.  Thereby the challenge is the construction of an appropriate subspace of the infinite Hilbert space of the electron and the phonons in the Hamiltionian~\eqref{H}. In this respect a very efficient approach is the variational ED (VED) \cite{BTB99,FT07,AFT10} method based on an increasing sequence of subspaces of the complete Hilbert space. It achieves an extreme accuracy for fermion-boson models with a finite number of particles with different types of couplings \cite{BT01,AEF07}  on an infinite lattice in any dimension. That means the method works for general polaron and bipolaron problems.

Here our starting state is a one-electron zero-phonon Bloch state with momentum $k$ in the  first Brillouin zone of an infinite chain, $|k \rangle \propto \sum_j e^{ikR{_j}} c^\dagger_j|0\rangle$, $|k|\le \pi $. Further basis states are generated by repeated 
action of the off-diagonal pieces of the Hamiltonian on this initial state. Taking advantage of the translational symmetry of~$\cal H$, 
for any new state generated (describing a new configuration of phonons relative to the electron) only one copy is retained.
One sees that VED gives the polaron properties as a continuous function of $k$, i.e., data are not only obtained at multiples 
of $
2\pi/N$  as for finite $N$-sites chains treated by standard ED.  

When we generate the basis set for the HM by applying the Hamiltonian $N_{\rm gen}$ times on the initial (zero phonon)  state, 
we obtain  a ``triangle" of phonon-containing states, consisting of a state with $N_{\rm gen}$ phonons at the electron site up to a state with a single phonon 
$(N_{\rm gen} -1)$ lattice sites away from the electron, and no phonon excitations elsewhere. For the EHM-3, where $j$ in $f_{j}(i)$ is
limited to the electron site and the both NN sites, after $N_{\rm gen}$ applications of ${\cal H}$ will have three 
states, one with $N_{\rm gen}$ phonons at the electron's initial site and two states with $N_{\rm gen}$ phonons  on NN sites. There will be also two states with one phonon at a distance  $N_{\rm gen}$  away from the  electron, and no phonons elsewhere. In that case this is the maximum distance 
from the electron site, that a phonon excitation can be found. 

At very strong EP coupling, where small polarons are formed, almost the entire lattice distortion is confined to the electron's site, and the Lang-Firsov (LF) transformed 
wave packet will be an appropriate starting  state \cite{LF62,LBBM10}:
\begin{equation}
|\psi (k)\rangle  = e^{-g^2 /2} \sum_{j} e^{ikR{_j}} e^{-g a_{j}^{\dagger}}|0 \rangle 
\end{equation}   
The corresponding LF-VED technique was described and tested in great detail in Refs.~\onlinecite{CMCD12,CM13,CTM14}.
In a nutshell, within the LF-VED we start from $(n+1)$ initial states, instead of just one zero-phonon state within standard VED.
In these $(n+1)$ states, we have up to $n$ phonon excitations  at the electron site, starting from 0. 
In the numerical work, we have been used $n$=$40$ for the HM and $n$=$30$ for EHM. For this we observed excellent convergence 
in the strong EP coupling limit, where the normal VED usually fails.

The largest variational basis used in this study takes $N_{\rm gen}=16$ for the LF-VED basis in the HM with NN hopping, which means $10\,857\,225$ configurations.
For the HM with NN and NNN hopping we include 8\,683\,122 configurations within VED ($N_{\rm gen}=17$) and 9\,398\,171 configurations within LF-VED using $N_{\rm gen}=12$ and $n$=$40$. For the EHM-3 with NN and NNN hopping the VED basis has 9\,592\,713 states  using $N_{\rm gen}=15$, whereas the LF-VED basis  has dimension 4\,963\,511     
for $N_{\rm gen}=10$ ($n$=$30$). For  EHM-5 with both kinds of hopping the VED basis size for $N_h=14$ is 9\,592\,713; for the corresponding EHM-7 the dimension is 8\,637\,123
when $N_{\rm gen}=13$. With this, for intermediate values of $g$, we get excellent convergence of the ground-state energies up to 11 digits.  Using the LF-VED in the strong EP coupling regime, we can maintain an accuracy of at least 6 digits.

\begin{figure}[t]
  \centering \includegraphics[width=0.7\columnwidth,angle=-90]{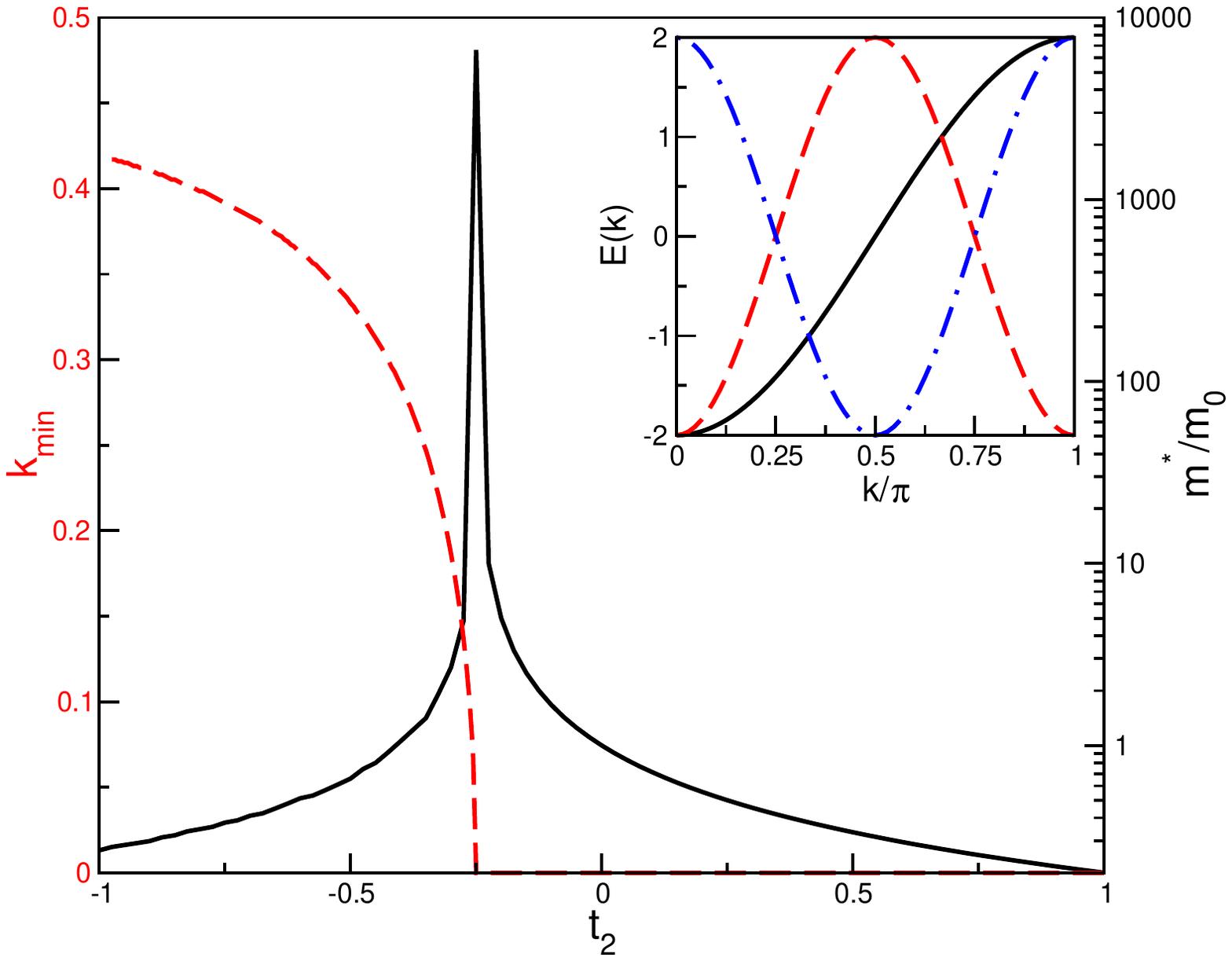}\\
  \caption{Momentum of the band minimum $k_{\rm min}$ of the bare electronic band dispersion~\eqref{T}  (red dashed line, left ordinate)  and effective band mass in units of $m_0=1/2t_1$ (black solid line, right ordinate)  as a function of the NNN hopping parameter $t_2$. The inset shows the limiting behavior of the bare band dispersion for $t_1=1$, $t_2=0$ (black solid line),  $t_1=0$, $t_2=1$ (red dashed line), and $t_1=0$, $t_2=-1$ (blue dot-dashed line).        
  \label{fig1}}
\end{figure}

\begin{figure*}[t]
  \centering \includegraphics[width=0.6\columnwidth,angle=-90]{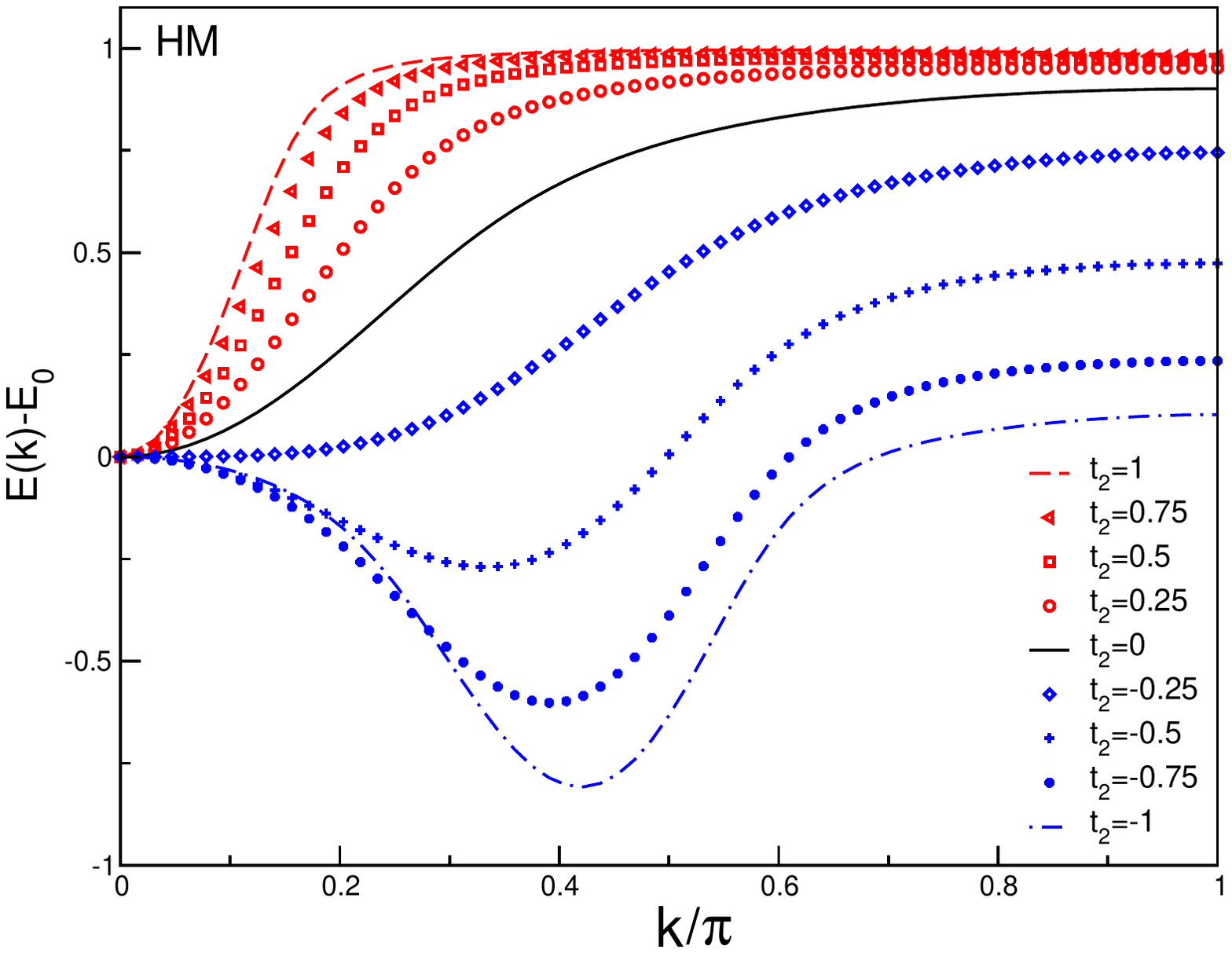}
  \centering \includegraphics[width=0.6\columnwidth,angle=-90]{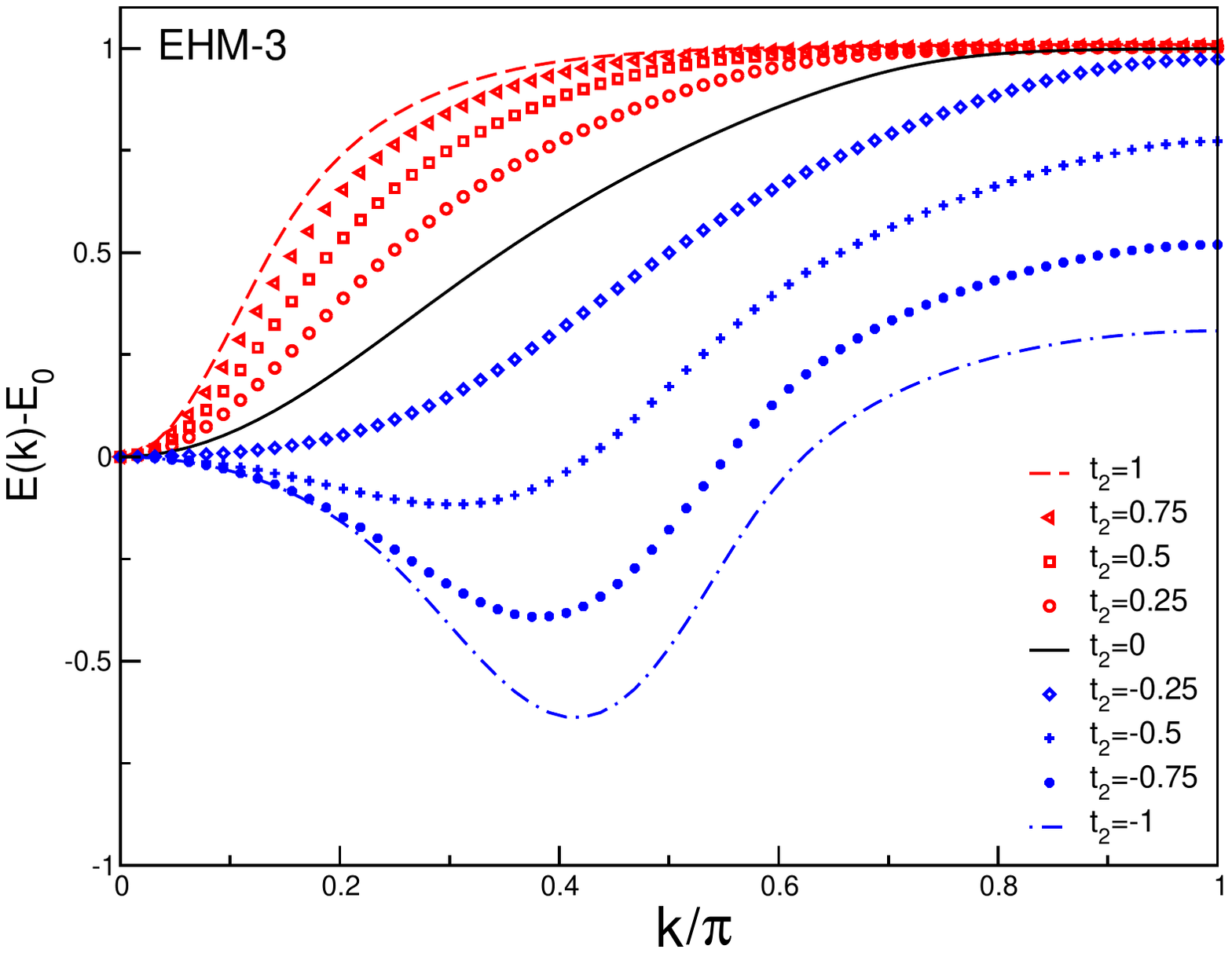}
  \centering \includegraphics[width=0.6\columnwidth,angle=-90]{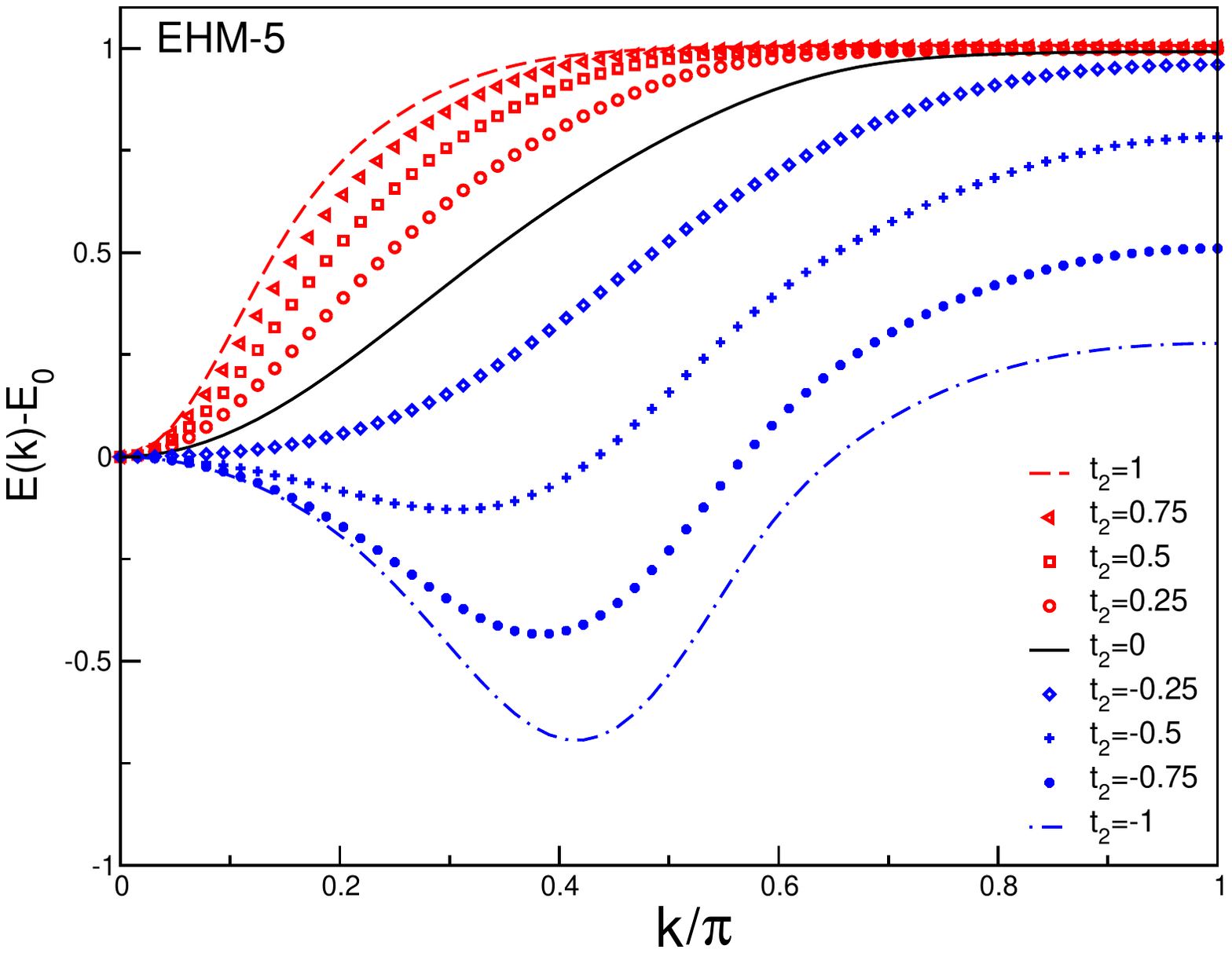}
  \centering \includegraphics[width=0.6\columnwidth,angle=-90]{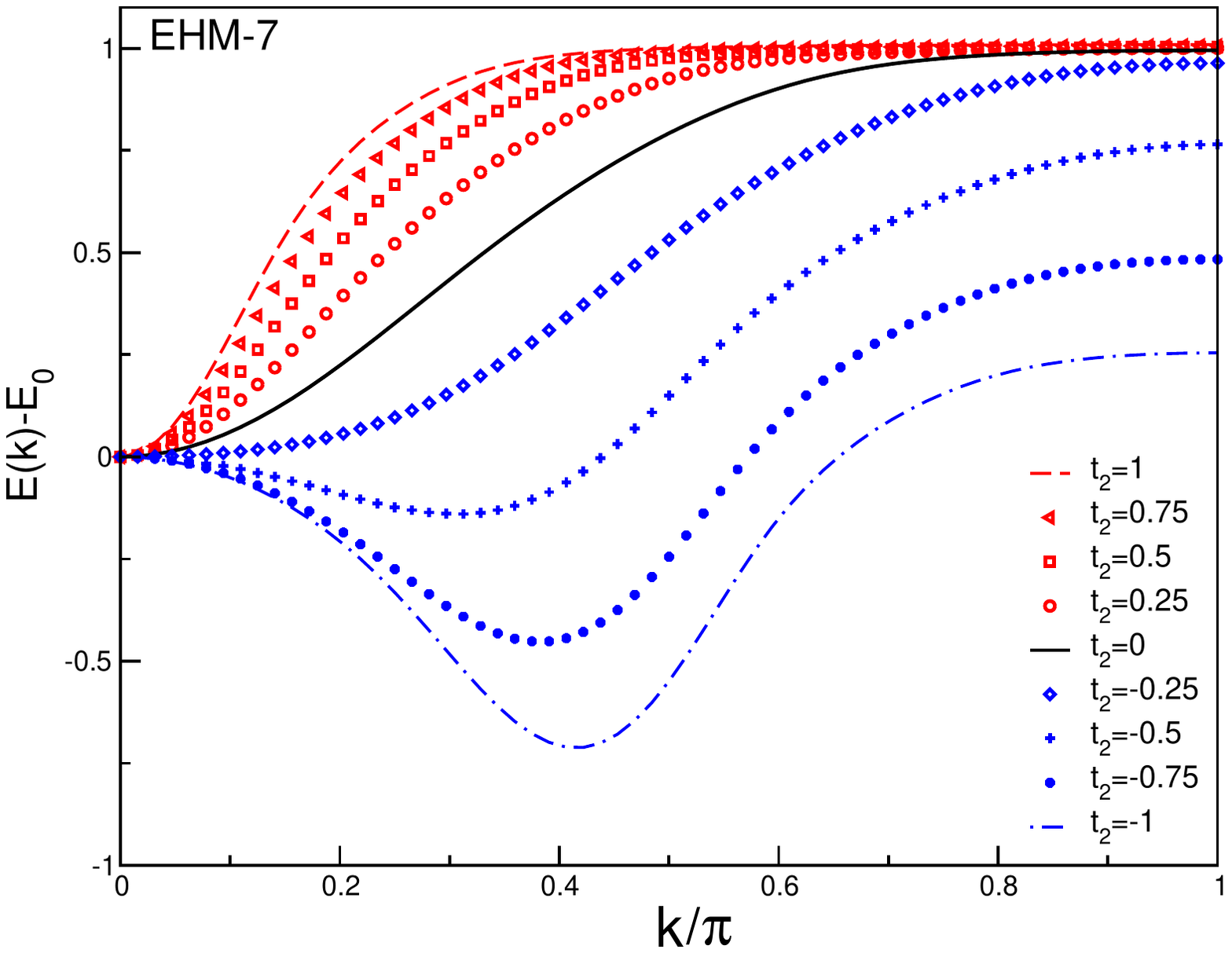}
\caption{Polaron band dispersion of the model~\eqref{H} for  different values of NNN hopping $t_2$. Shown are data for the HM (top left), EHM-3 (top right), EHM-5 (bottom left) and EHM-7 (bottom right). In all panels the EP coupling $g$=$1$  and bare phonon  frequency $\omega=1$. $E_0$ denotes the homogeneous band lowering caused by the EP interaction.
\label{fig2}}
\end{figure*}
\begin{figure}[t]
\centering \includegraphics[width=0.7\columnwidth,angle=-90]{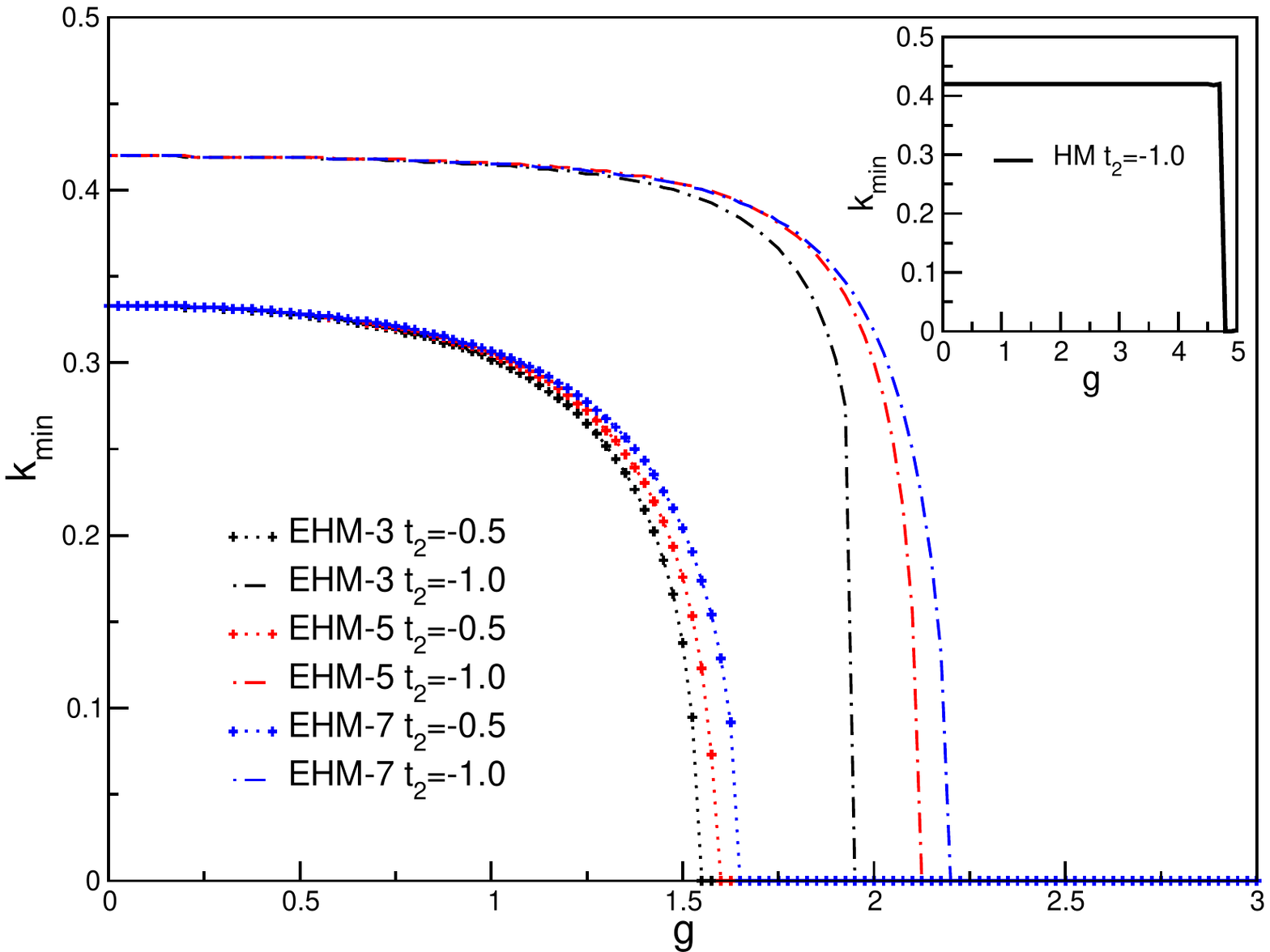}
\caption{ Location of the band minimum $k_{\rm min}$ in the EHM with $\omega=1$ as a function of $g$ for  different ranges of the EP interaction in EHM-3, EHM-5 and EHM-7. 
The inset gives the variation of $k_{\rm min}$ in the HM  with $t_2=-1$ as a function of the EP coupling strength $g$.
\label{fig3}}
\end{figure}

\begin{figure}[t]
\centering \includegraphics[width=0.7\columnwidth,angle=-90]{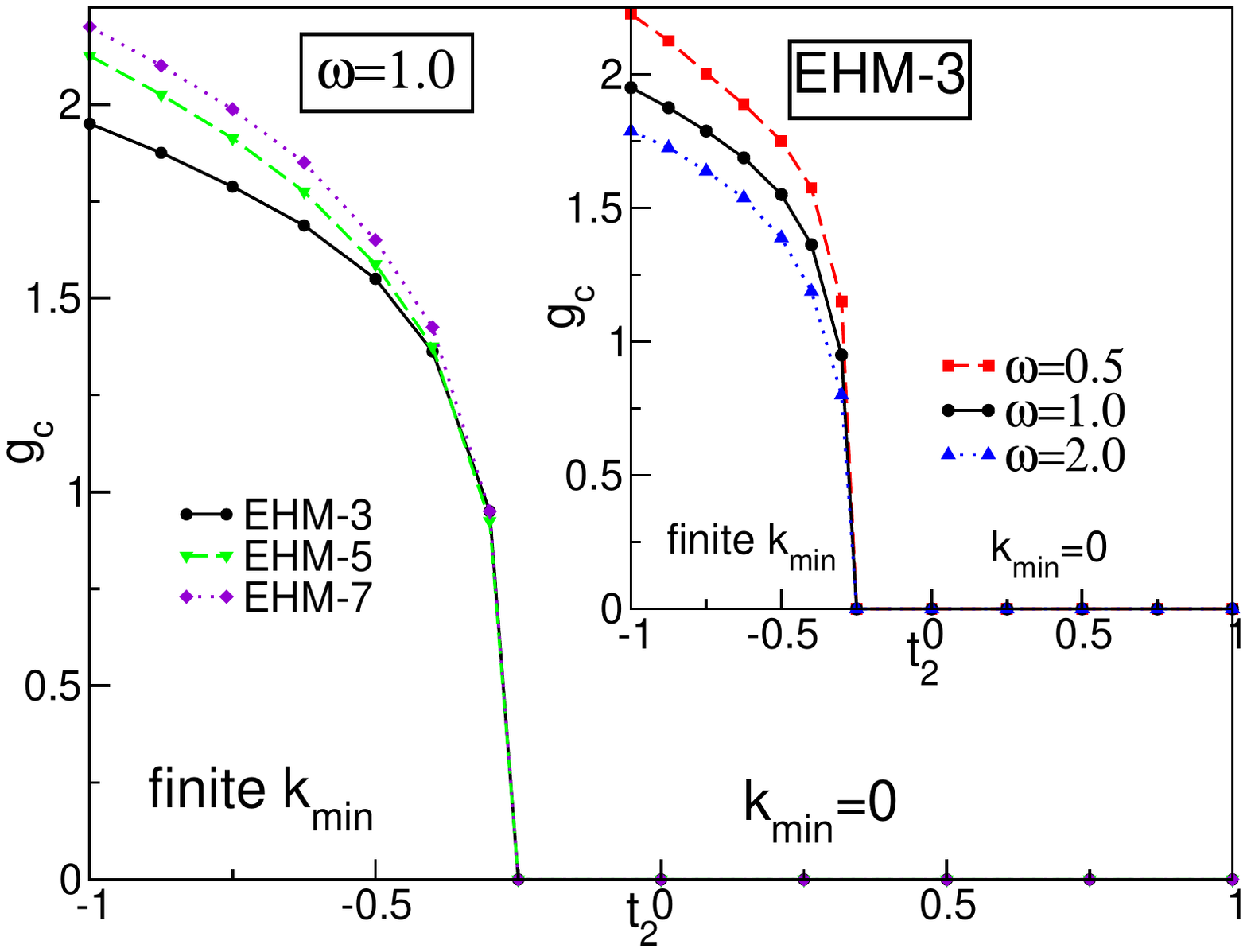}
\caption{ Ground-state  regions of the EHM with NN and NNN transfer having $k_{\rm min}=0$ and a finite $k_{\rm min}$. The main panel compares the phase boundaries of EHM-3, EHM-5 and EHM-7 at $\omega$=$1.0$. The inset shows the phase diagram of the EHM-3 for three different phonon frequencies.}
 \label{fig_PhaseDia}
\end{figure}

\section{Numerical Results and Discussion}
To set the stage for a discussion of the EP coupling effects in the full (E)HM Hamiltonian ${\cal H}$, let us first consider the free electron transfer ${\cal T}$ given by~\eqref{T}.  The corresponding band dispersion in momentum space is $E(k)=-2t_{1}cos(ka) -2t_{2}cos(2ka)$. If $t_2=0$, of course, we end up with the usual tight-binding (hyper-cubic) band structure, having its minimum at $k=0$ in the first Brillouin zone, provided that $t_1>0$. The situation changes including the NNN hopping term ($\propto t_2$). 
Now for  $t_2/t_1<-0.25$ the band minimum will be at a finite momentum $k_{\rm min}>0$  (let us consider positive momenta only). 
Interestingly at $t_2/t_1=-0.25$ the band curvature  vanishes; i.e., the particle mass 
\begin{equation}
m^*= \left[ \frac{\partial^2E(\vec{k})}{\partial {\vec{k}}^2} \right]_{k=k_{\rm min}}^{(-1)}
\label{MEFF}
\end{equation}
diverges. This is illustrated by Fig.~\ref{fig1}, where $m^*$ has been measured in units of the bare NN hopping band mass $m_0=1/2t_1$. We wish to stress that this mass enhancement is of purely electronic (band structure) origin. 

Quite naturally the particle's interaction with the lattice degrees of freedom leads to a renormalization of the band structure \cite{St96,WF97,FLW97,FLW00,CDC11,CPM16,CTB17}. Figure~\ref{fig2} displays the polaron band dispersion of the HM and EHM for different $t_2/t_1$ ratios in the intermediate EP coupling and phonon frequency regime ($g=\omega=1$), which is most difficult to address by analytical approaches. Results for EHM are given for different ranges of the EP coupling (involving 3, 5, and 7 neighboring sites).  Thereby, $E(k)$ denotes the so-called ``coherent" band dispersion obtained, within (V)ED approaches, by the lowest energy value in each $k$ sector. As a rule, $E(k)$ conforms with the first peak in the single-particle spectral function \cite{FLW00}. For $t_2\geq0$, the band minimum is always at the band center ($k=0$) and,  because $g$ is not that big, the dispersion is only slightly renormalized at small momentum $k$.  The situation is quite different at larger momenta: Here a band ``flattening'' is observed when the dispersionsless phonon branch intersects the electronic band structure. We would expressly like to emphasize that this effect appears at weak-to-intermediate EP coupling and for small-to-moderate phonon frequencies only, and is most pronounced near the zone boundary \cite{WF97,FLW97,FLW00}. In the very strong EP coupling regime, the bandwidth of the polaron band substantially reduces to values much less than the bare phonon frequency, which means that this kind of flattening'' is negligible. For $t_2 < 0$, the bandwidth becomes smaller. For the HM it reaches its minimal value in the range of $t_2=-0.25$, where the band curvature at $k=0$ almost disappears; afterward the bandwidth increases again when $t_2$ is getting smaller.  As stated above, in this region  the band structure develops a minimum at finite momentum $k_{\rm min}$.  The influence of the range of the EP interaction on this general behavior is  very minor, as can be seen by comparing the EHM-3, EHM-5, and EHM-7 data.

Interestingly, we can induce a sharp transition from a polaronic state with finite momentum to one with zero momentum by increasing the EP coupling strength at otherwise fixed parameters. To demonstrate this striking feature, we show in  Fig.~\ref{fig3} how $k_{\rm min}$ varies when $g$ is raised. Obviously,  a critical EP coupling $g_c$ exists for the (E)HM with large negative $t_2/t_1$ ratios. Note that the value of $g_c$ increases when decreasing $t_2$ and including longer-ranged EP interactions. The inset of Fig.~\ref{fig3} gives $k_{\rm min}$ as a function of $g$ for the HM. Again we find a sharp transition, but at too high $g$  to be of any physical relevance.

Figure~\ref{fig_PhaseDia} gives a kind of ground-state phase diagram of the EHM with NNN transfer in the $t_2-g_c$--plane, where the lines separate ground states with $k_{\rm min}=0$ and finite $k_{\rm min}$. 
The inset demonstrates for the EHM-3 that (i) the $k_{\rm min}$-finite region is bounded by $t_2/t_1=-0.25$ (at $g=0$) and (ii) the sharp polaron transition to the $k_{\rm min}=0$ ground state appears at smaller values of the EP coupling when going from the rather adiabatic ($\omega=0.5$) to the non-adiabatic ($\omega=2$) phonon-frequency regime, i.e., in the adiabatic case, where more energetically less costly phonons are involved, stronger EP coupling is necessary to renormalize the band structure for $t_2/t_1\leq -0.25$. Most notably, however, the general quantum phase transition behavior of the EHM with NNN hopping does not depend on the value of the phonon frequency. This is underlined by the main panel comparing the ground-state momentum regions of EHM-3, EHM-5 and EHM-7. Here, a longer-ranged EP coupling clearly leads to a larger critical coupling $g_c$.

\begin{table*}[t]
\caption {Effective transfer integrals $\tilde{t}_1$, $\tilde{t}_2$, $\tilde{t}_3$ and $\tilde{t}_4$
for the EHM-3  with  $t_2=-1$ and $\omega=1$ at various  $g$. }
\begin{ruledtabular}
\begin{tabular}{|c|c|c|c|c|}
$g$&  $\tilde{t}_1$&$\tilde{t}_2$&$\tilde{t}_3$ &$\tilde{t}_4$ \\  \hline
 0.0& 1.0 &-1.0 &0.00 &0.000  \\ \hline
 0.25 &0.0662707& -0.17598 & -0.129112  & 0.0591186\\ \hline
 0.5 &0.0810571& -0.177426 & -0.12489  & 0.0507156\\ \hline
 1.0 &0.129658& -0.150741 &-0.0769285  &0.0307539\\ \hline
 1.5 &0.121126 &-0.0751145 &-0.0237463 &0.0093872   \\ \hline
 2.0 & 0.071929 &-0.0172384 &-0.0033003 &0.00095433   \\ \hline
\end{tabular}
\end{ruledtabular}
\end{table*}

In order to analyze more thoroughly the impact of the EP coupling 
 on the shape of the polaron band as a whole and the just discussed abrupt transition in particular, we have 
fitted the polaron band dispersion to the following functional form, 
\begin{equation}
E(k) = -  \sum_{n=0}^{n_{\rm max}} 2\tilde{t}_{n} cos(nk)\,,
\label{Fit}
\end{equation}   
including direct effective hopping between lattice sites that are up to $n_{\rm max}$ places away from each other \cite{MSB17}. 

Table I gives the fitting parameters $\tilde{t}_n$ for different EP couplings $g$ at $\omega=1$, where we have considered the most extreme negative $t_2=-1$ band case. The numbers yield the following trends. First, the EP coupling induces longer-ranged particle hopping processes (beyond those  appearing in the Hamiltonian), especially in the intermediate coupling and frequency regime. For example, in order to get a satisfactory fit at $g=0.25$ and $g=0.5$ an $n_{\rm max}$ of the order of $20$ was necessary (note that Table~I give the first four hopping parameters only). Second, in the very strong EP coupling limit, where the band minimum always occurs at $k=0$ and the bandwidth is reduced dramatically, {\it de facto} only the very small effective NN transfer matters, irrespective of the range of both the bare direct electron hopping and of the EP interaction.  Third, the magnitudes of the different effective transfer integrals significantly change when passing the critical EP coupling $g_c$, signaling the finite to zero momentum transition. 

\begin{figure}[t]
  \centering \includegraphics[width=0.7\columnwidth,angle=-90]{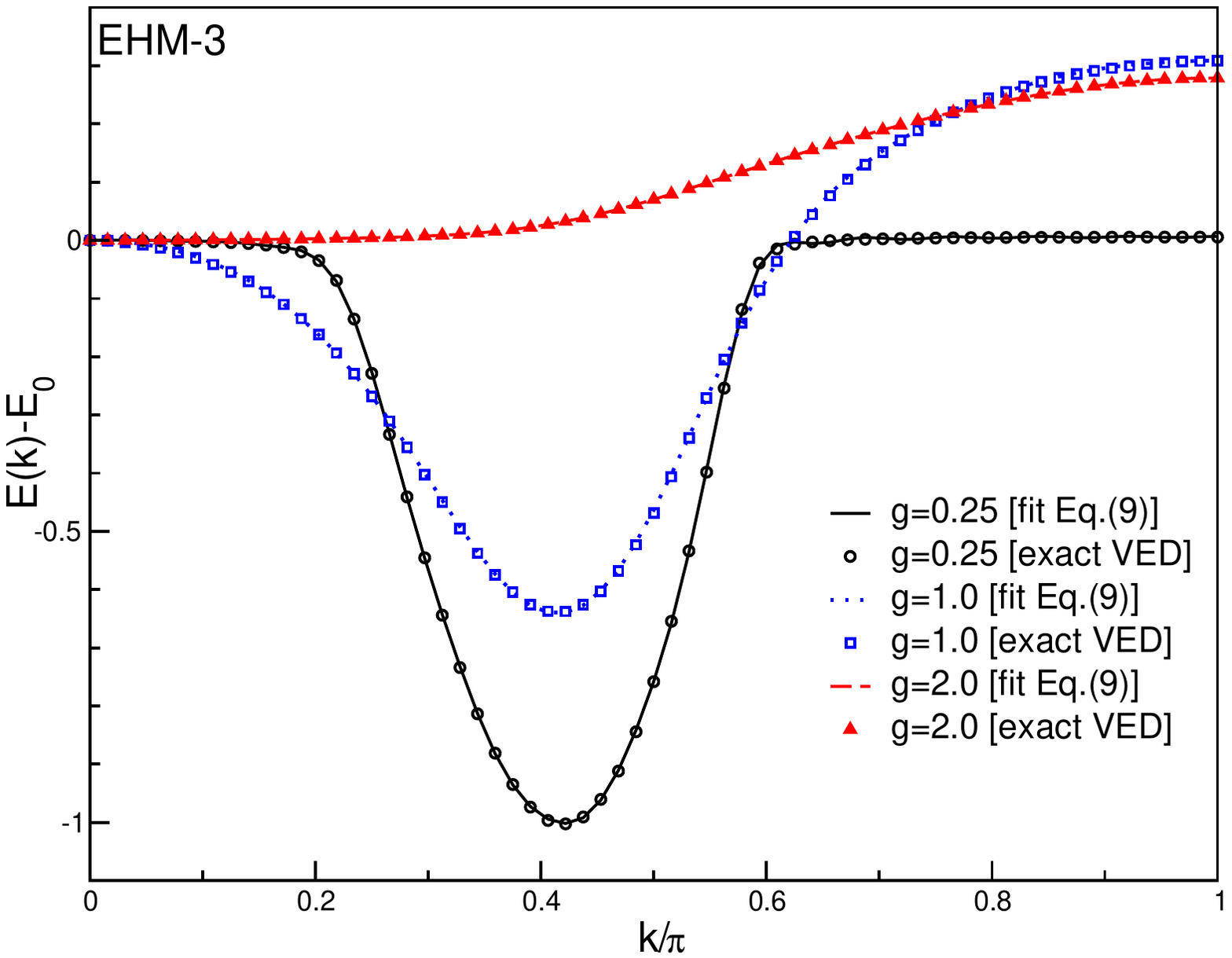}
  \caption{Band dispersion of the EHM-3 with NN and NNN hopping $t_2/t_1=-1$ for different EP couplings $g$ at phonon frequency $\omega=1$.
  Exact VED data (symbols) are compared with the $E(k)$ fit according to Eq.~\eqref{Fit} with  $n_{\rm max}= 20$  for $g=0.25$ and  $n_{\rm max}= 9$ for $g=1.0, 2.0$. 
  \label{fig4fit}}
\end{figure}

To demonstrate the accuracy of our band-dispersion fitting procedure, we compare   $E(k)$ with the exact VED results for $t_2/t_1=-1$ in Fig.~\ref{fig4fit}. Note that we obtain already an excellent fit for $n_{\rm max}=9$, provided that $g\gtrsim 1$, whereas an  $n_{\rm max}=20$  is necessary to get sufficiently accurate results at a smaller value $g\simeq 0.25$.

\begin{figure}[t]
  \centering \includegraphics[width=0.7\columnwidth,angle=-90]{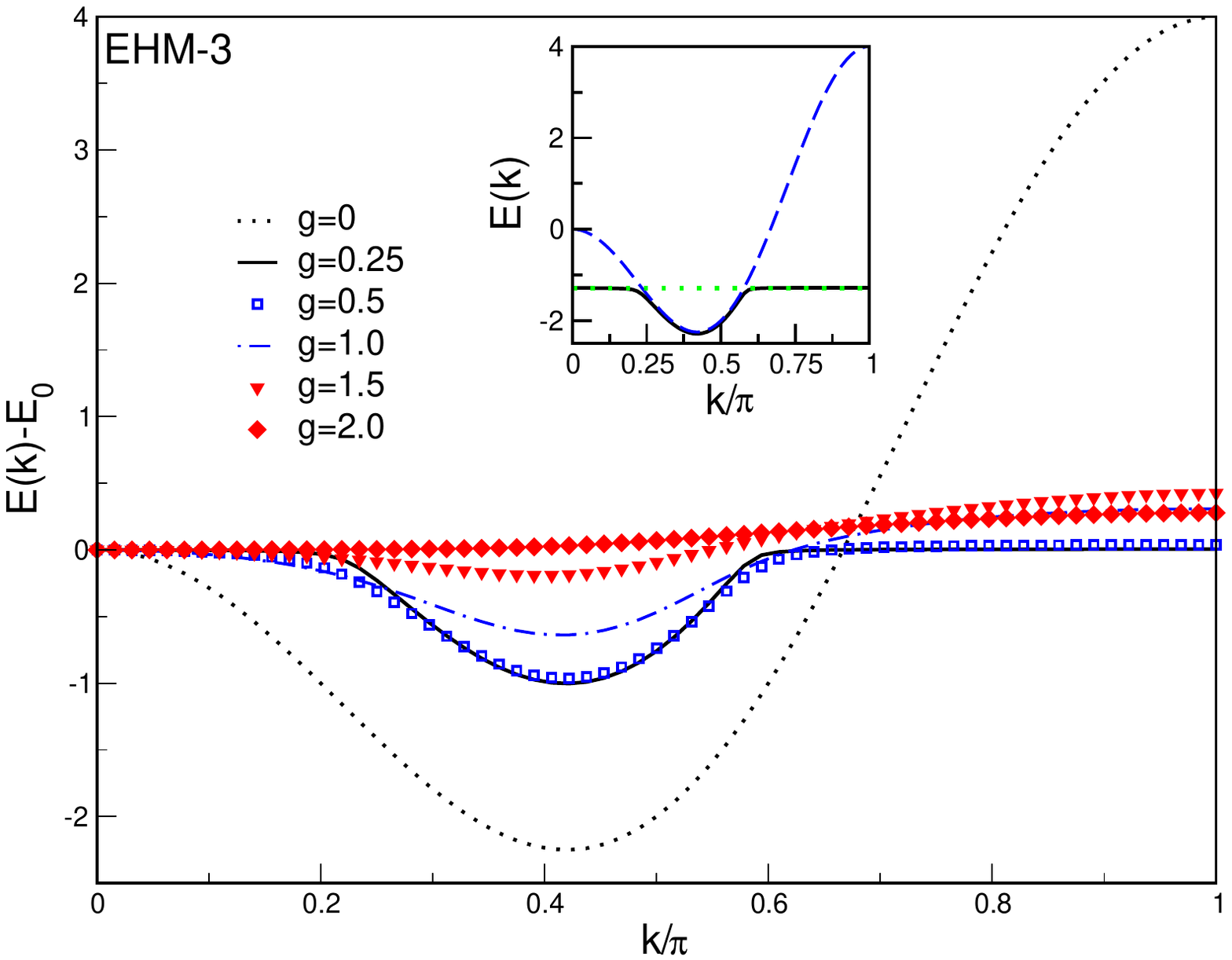}
  \caption{Polaron band dispersion  for the EHM-3 with NN and NNN hopping with $t_2/t_1=-1$ for different EP couplings $g$ at phonon frequency $\omega=1$. The inset shows the bare electronic band for $t_2/t_1=-1$ (blue dashed curve) intersected by the bare phonon [horizontal green line located at $E(k)_{\rm min}+\omega$]. The black solid line shows the resulting polaronic band $E(k)$ at $g=0.25$.        
  \label{fig4}}
\end{figure}

Figure~\ref{fig4} presents the development of the EHM-3 polaron band with increasing EP coupling at $t_2=-1$ (again $\omega=1$).  Starting from the non-interacting band structure with $k_{\rm min}\simeq 0.4)$ for $g$ (black dotted line), we first can observe a stronger band narrowing by raising $g$ from 0.25 to 1 (dashed dotted line) together with an pronounced  band flattening away from momentum  $k_{\rm min}$ by  virtue of the intersecting phonon branch (cf. also the inset). Then, in the range between $g=1.5$ to 2.0, the transition to a  
strongly renormalized polaron band takes place, which is accompanied by the shift of the momentum of the ground state to $k=0$. We note that this scenario is just opposite to what happens in the SSH model, where, if the EP coupling becomes larger, a noticeable effective NNN hopping of negative sign is created that finally shifts the momentum of the ground state from zero to finite momentum.

\begin{figure}[t]
\centering \includegraphics[width=0.8\columnwidth,angle=-90]{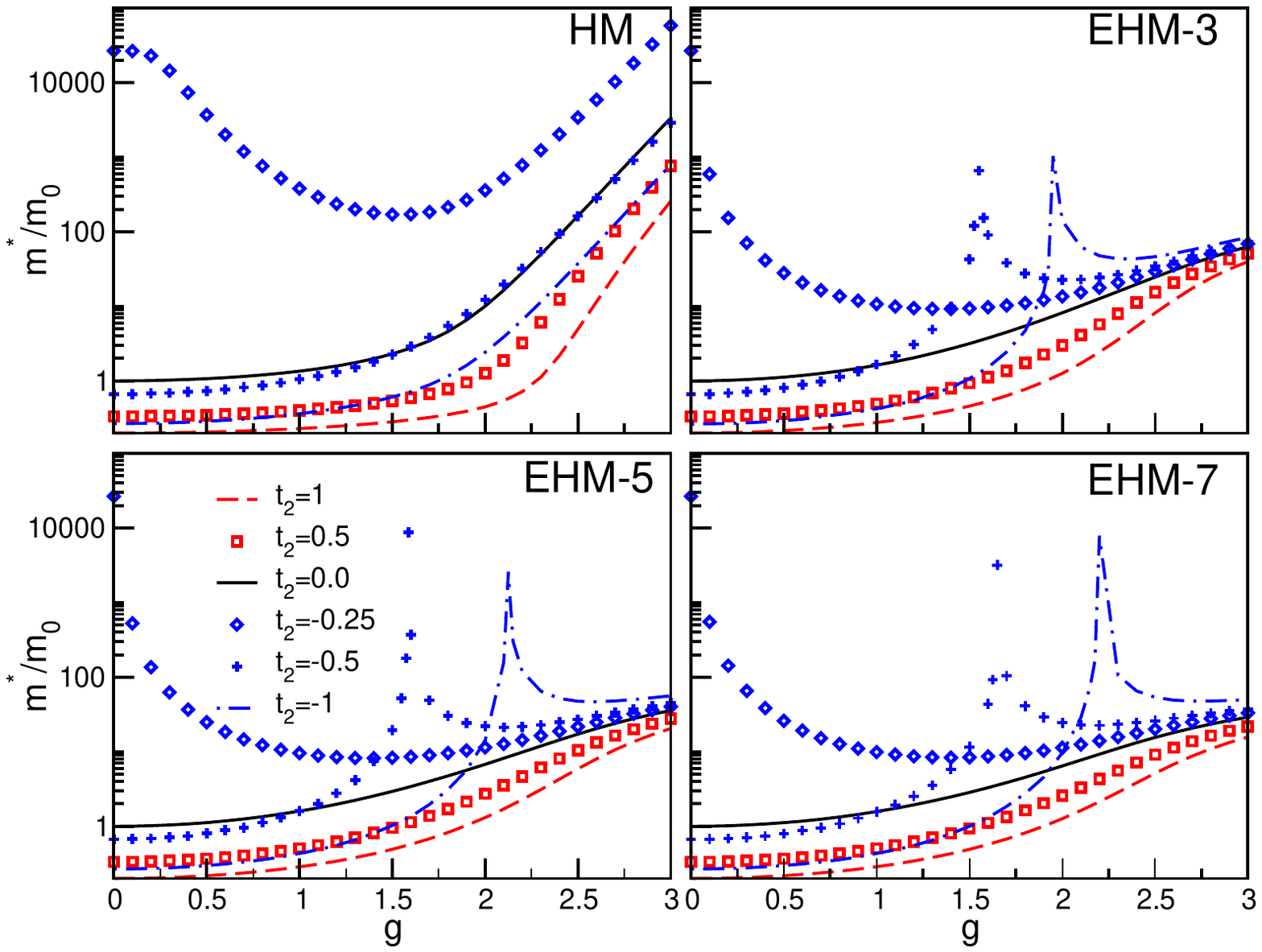}
\caption{ Effective mass for the HM and EHM with $\omega$=$1$ as a function of $g$ at various NNN hoppings $t_2=1$ (dashed red line), $t_2=0.5$ (red squares), $t_2=0$  (black solid line), $t_2=-0.25$ (blue diamonds), $t_2=-0.5$ (blue crosses),  $t_2=-1$ (blue dot-dashed line). 
 \label{fig5}}
\end{figure}

In the discussion of polaron formation the perhaps most noteworthy quantity is the effective mass. Increasing the EP coupling the polaron mass usually grows; i.e., the mass renormalization $m^*/m_0$ becomes larger than unity. Because the (LF-)VED method works on the infinite lattice with continuous momentum $k$ we can directly determine 
the second $k$-derivative needed in Eq.~\eqref{MEFF}. We should point out, however, that although the calculated energies $E(k)$ are variational bounds for the corresponding exact energies, there is no such bound for the effective mass that might be either above or below the exact value. Figure~\ref{fig5} gives the values of the effective mass as a function of the EP coupling strength $g$ for the (E)HM with different NNN hopping $t_2$.  Considering the HM data first, we observe the general increase of $m^*/m_0$ with $g$ mentioned above, provided that the ratio $t_2/t_1$ is  away from the critical one $t_2/t_1 =- 0.25$. In the latter case, a huge mass enhancement appears even at $g=0$, which is a purely electronic band-structure effect.  Consequently, increasing the EP coupling,  $m^*/m_0$ decreases first before it increases again at larger values of $g$. Furthermore, we see that the direct NNN transfer tends to decrease the effective mass, in particular if the ratio $t_2/t_1$ is positive. Compared to the HM, the mass enhancement in the EHM is weaker \cite{AK99,FLW00}. Obviously, the larger spatial extension of the phonon cloud surrounding the particle supports its mobility. Most notably, we observe a pronounced cusp in 
$m^*/m_0$ when passing the band-structure transition from finite to zero momentum by increasing the EP coupling at negative $t_2/t_1$.

\begin{figure}[t]
  \centering \includegraphics[width=0.7\columnwidth,angle=-90]{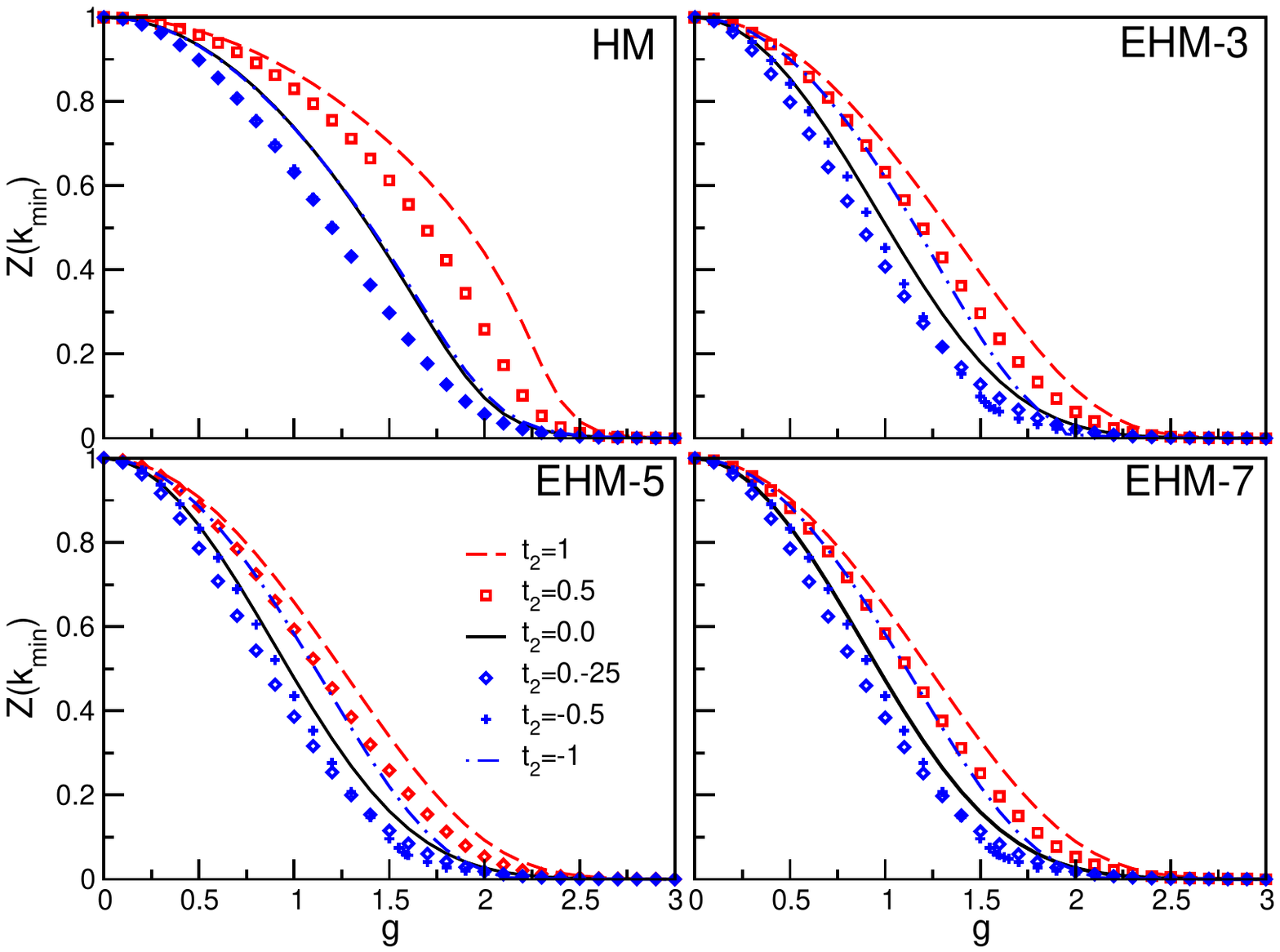}
  \caption{Wave-function renormalization factor  (quasiparticle weight) for the HM, EHM-3, EHM-5, and EHM-7 with $\omega$=$1$  as a function of the EP-coupling strength $g$. Notations are the same as in Fig.~\ref{fig5}.
  \label{fig6}}
\end{figure}

In order to gain deeper  insight into the nature of polaronic band states we take a look at the 
wave-function renormalization factor,   
\begin{equation}
Z(k)=\vert\langle \psi_k \vert c_{k}^{\dag} \vert 0\rangle\vert^{2}\,,
\label{Z}
\end{equation}
where $| \psi_k \rangle$  denotes the polaron state with momentum $k$ being lowest in energy and $\vert 0\rangle$ represents the vacuum state.  Hence, $Z({k})$ gives the electronic spectral weight of the first peak in the wave-vector-resolved single-particle spectral function with momentum $k$, and $Z(k_{\rm min})$ is referred to as ``quasiparticle weight''. Figure~\ref{fig6} shows the variation of 
$Z(k_{\rm min})$ when increasing the EP coupling in the different model Hamiltonians. Clearly we have $Z(k_{\rm min})=1$ at $g=0$ (free electron case). For $g>0$,  $Z(k_{\rm min})<1$ can be taken as a measure of  how much the polaron deviates from the free electron. Here any significant reduction of $Z(k_{\rm min})$  signals a strong dressing of the electron by a phonon cloud.  $Z(k_{\rm min})\ll 1$  in the very-strong EP regime where heavy polarons emerge.  
At a given value of the EP coupling, the quasiparticle weight is smallest (largest) for $t_2/t_1=-0.25$ ($t_2/t_1=1$) where the electronic mobility  is  smallest 
(largest), making the EP interaction that tends to ``trap'' the charge carrier by forming a polaron more effective (ineffective). A longer-ranged EP coupling enhances the extent 
of the phonon cloud and thereby reduces the electronic component of the quasiparticle [see the reduction of  $Z(k_{\rm min})$ when comparing the HM and EHM data]. Lastly we would like to point out, that electronic wave-function renormalization factor $Z(k)$ for momenta in the flat-band regions is also considerably suppressed, simply because these states are rather phononic than electronic in nature.

\begin{figure}[t]
  \centering \includegraphics[width=0.7\columnwidth,angle=-90]{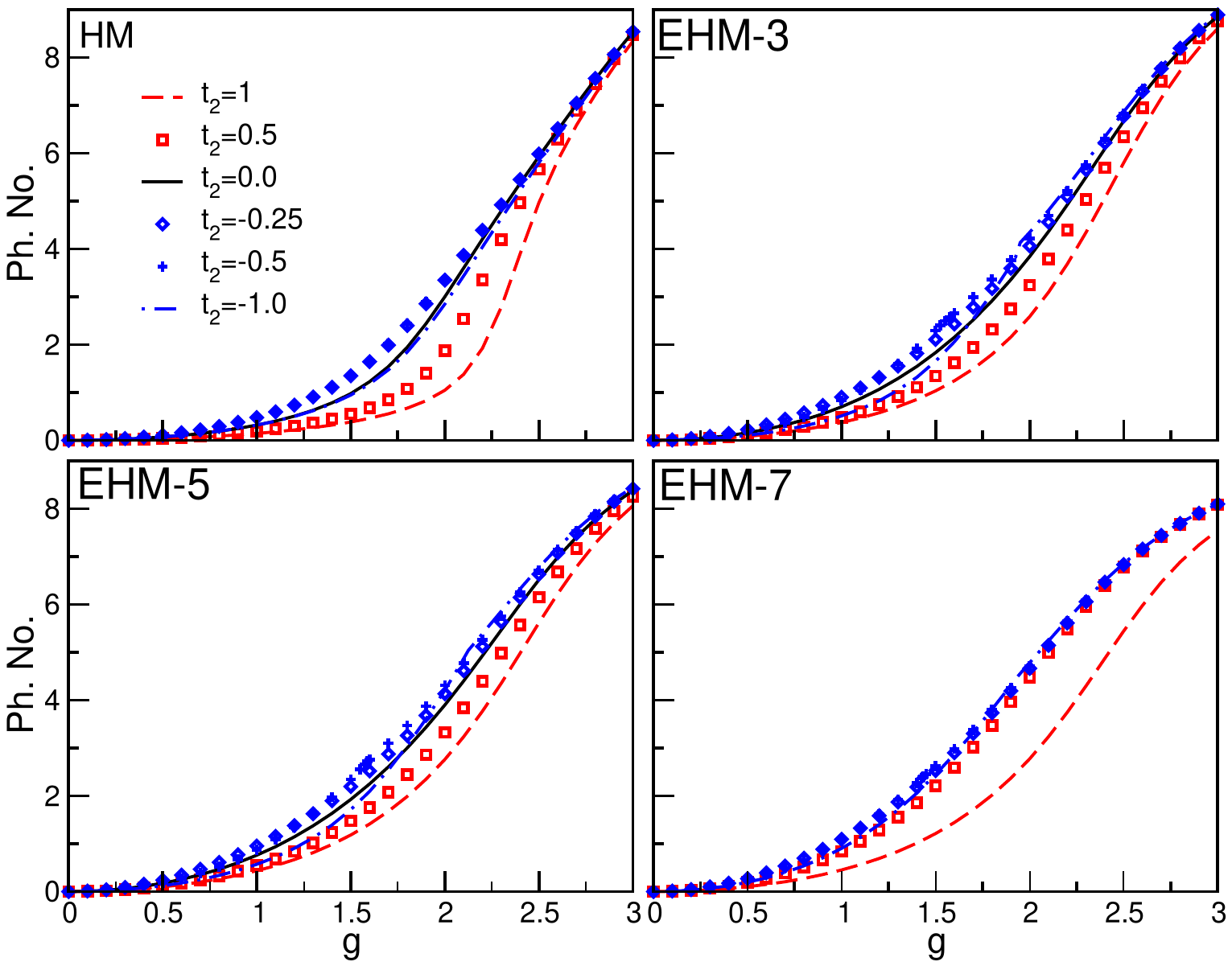}   
\vskip 0.1 cm 
\caption{Average phonon number  in the ground state of the HM and EHM with $\omega$=$1$ as a function of $g$. Notations are the same as
in Fig.~\ref{fig5}.  
  \label{fig7}}
\end{figure}

This scenario is corroborated when tracking the average phonon number in the ground state for the same model parameters see (Fig.~\ref{fig7}). As expected the number of phonons is small (large) for weak (strong) EP coupling, and the tendency with respect to a variation of the range of the coupling and the sign and the magnitude NNN hopping   is consistent with what has been discussed for $Z(k_{\rm min})$ above.

\begin{figure}[t]
\centering \includegraphics[width=0.7\columnwidth,angle=-90]{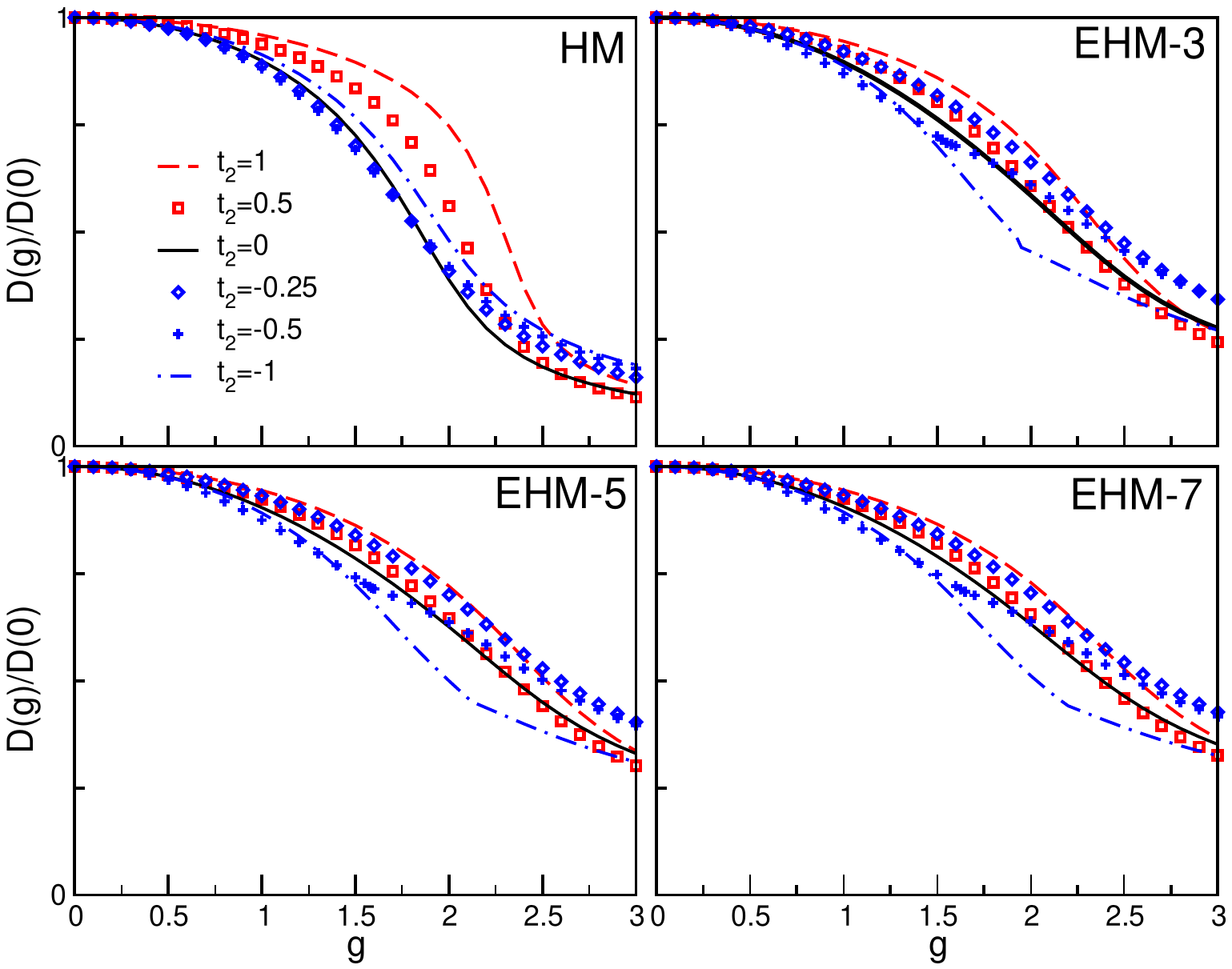}
\caption{ Drude weight for the HM and EHM with $\omega$=$1$ as a function of $g$. Notations are the same as
in Fig.~\ref{fig5}.
 \label{fig8}}
\end{figure}

The VED scheme can also be used to investigate the transport and optical response of polaronic systems \cite{FT07}. Here we will focus on the so-called Drude
weight only (see Fig.~\ref{fig8}), which is a measure of  ``coherent'' free-particle-like transport with zero energy transfer.  For the (E)HM the Drude weight can be calculated from Kohn's formula ~\cite{Ko64},  
\begin{equation}
D=\left. \frac{\partial^2E_0(\phi)}{\partial \phi^2} \right|_{\phi=0}
\label{D}
\end{equation}
(in units of $\pi e^2$), where $E_0(\phi)$ is the ground-state energy of our system in the presence of a non-vanishing phase factor $\phi$, which is introduced in both the $t_1$ and $t_2$ transfer amplitudes. This breaks the time-reversal symmetry of ${\cal H}$. Not surprisingly, the Drude weight  is getting suppressed enhancing the EP interaction and again this effect is more pronounced in the HM in comparison to the EHM. Most notable is the kink appearing in the $t_2$=$-1$ and $t_2$=$-0.5$ curves at the critical EP coupling for the finite-to-zero momentum transition. This behavior is in accordance with the change in the relative strength of the different effective hopping integrals listed in Table I and should be detectable in transport measurements.

\section{Conclusions}
Lattice polaron formation strongly depends on the kind, range and strength of the EP interaction, the (different) times scales of electron and phonon dynamics, as well as on the (bare) electronic band structure and the spatial dimension of the system. As a result of this interplay very complex and nontrivial quantum correlations develop. 
These determine not only the properties of the polaronic quasiparticle itself, but also the nature of the polaron transition. The (single) polaron transition 
is a continuous crossover; at least the ground state is an analytic function of the EP coupling at finite phonon frequency in any dimension \cite{GL91,AFT10}. This is believed to be 
the case for standard Holstein- and Fröhlich-type models. If the EP coupling is nondiagonal in the site index, as in the SSH model, a critical EP coupling exists, 
at which a sharp transition between states with zero and finite momentum takes place, albeit the process is put in reverse compared to the SSH model \cite{MDCBNPMS10}. The reason is the well-known effective long-ranged particle hopping 
processes induced by the EP coupling dynamically. Such processes will be also induced in the (extended) Holstein models \cite{St96,WF97} and might interfere with longer-ranged electronic hopping terms in the (bare) band structure if present.  In this context, we have studied the extended Holstein model with nearest and next nearest neighbor transfer 
and proved numerically, by means of an unbiased variational exact diagonalization method, that a sharp transition related to a polaron ground-state momentum change might also exist, even at finite phonon frequencies. This transition, appearing for $t_2/t_1<-0.25$, is mainly triggered by the ratio between the effective NN and NNN transfer intergrals (which are renormalized by the EP coupling of course) and accompanied by a dramatic mass (mobility) increase (decrease), whereas the particle's phonon dressing, wave-function renormalization factor and Drude weight  are less affected. Away from this transition, compared to the HM polaron, the EHM polaron has a larger extent and the polaron band is less renormalized. At the same time, the number of  phonons the (EHM) charge carrier has to drag through the lattice when coherently moving is larger than those for the HM polaron. We further note that while the inverse polaron effective  mass is directly given by the wave-function renormalization factor for the HM with NN hopping, this relation is much more complicated for the (E)HM with NNN transfer. This shows the complex interplay between band structure and EP coupling effects in these types of models.\\

\acknowledgements  M.C. appreciates access to the computing facilities of the DST-FIST (phase-II) project installed in the Department of Physics, IIT Kharagpur, India. 

\begin{thebibliography}{68}%
\makeatletter
\providecommand \@ifxundefined [1]{%
 \@ifx{#1\undefined}
}%
\providecommand \@ifnum [1]{%
 \ifnum #1\expandafter \@firstoftwo
 \else \expandafter \@secondoftwo
 \fi
}%
\providecommand \@ifx [1]{%
 \ifx #1\expandafter \@firstoftwo
 \else \expandafter \@secondoftwo
 \fi
}%
\providecommand \natexlab [1]{#1}%
\providecommand \enquote  [1]{``#1''}%
\providecommand \bibnamefont  [1]{#1}%
\providecommand \bibfnamefont [1]{#1}%
\providecommand \citenamefont [1]{#1}%
\providecommand \href@noop [0]{\@secondoftwo}%
\providecommand \href [0]{\begingroup \@sanitize@url \@href}%
\providecommand \@href[1]{\@@startlink{#1}\@@href}%
\providecommand \@@href[1]{\endgroup#1\@@endlink}%
\providecommand \@sanitize@url [0]{\catcode `\\12\catcode `\$12\catcode
  `\&12\catcode `\#12\catcode `\^12\catcode `\_12\catcode `\%12\relax}%
\providecommand \@@startlink[1]{}%
\providecommand \@@endlink[0]{}%
\providecommand \url  [0]{\begingroup\@sanitize@url \@url }%
\providecommand \@url [1]{\endgroup\@href {#1}{\urlprefix }}%
\providecommand \urlprefix  [0]{URL }%
\providecommand \Eprint [0]{\href }%
\providecommand \doibase [0]{http://dx.doi.org/}%
\providecommand \selectlanguage [0]{\@gobble}%
\providecommand \bibinfo  [0]{\@secondoftwo}%
\providecommand \bibfield  [0]{\@secondoftwo}%
\providecommand \translation [1]{[#1]}%
\providecommand \BibitemOpen [0]{}%
\providecommand \bibitemStop [0]{}%
\providecommand \bibitemNoStop [0]{.\EOS\space}%
\providecommand \EOS [0]{\spacefactor3000\relax}%
\providecommand \BibitemShut  [1]{\csname bibitem#1\endcsname}%
\let\auto@bib@innerbib\@empty
\bibitem [{\citenamefont {Landau}(1933)}]{LA33}%
  \BibitemOpen
  \bibfield  {author} {\bibinfo {author} {\bibfnamefont {L.~D.}\ \bibnamefont
  {Landau}},\ }\href@noop {} {\bibfield  {journal} {\bibinfo  {journal} {Phys.
  Z. Sowjetunion}\ }\textbf {\bibinfo {volume} {3}},\ \bibinfo {pages} {664}
  (\bibinfo {year} {1933})}\BibitemShut {NoStop}%
\bibitem [{\citenamefont {Mott}\ and\ \citenamefont {Gurney}(2014)}]{MG40}%
  \BibitemOpen
  \bibinfo {editor} {\bibfnamefont {N.~F.}\ \bibnamefont {Mott}}\ and\ \bibinfo
  {editor} {\bibfnamefont {R.~W.}\ \bibnamefont {Gurney}},\ eds.,\ \href@noop
  {} {\emph {\bibinfo {title} {Electronic Processes in Ionic Crystals}}},\
  \bibinfo {edition} {eighth edition}\ ed.\ (\bibinfo  {publisher} {Clarendon
  Press},\ \bibinfo {address} {Boston},\ \bibinfo {year} {2014})\BibitemShut
  {NoStop}%
\bibitem [{\citenamefont {Firsov}(1975)}]{Fi75}%
  \BibitemOpen
  \bibfield  {author} {\bibinfo {author} {\bibfnamefont {Y.~A.}\ \bibnamefont
  {Firsov}},\ }\href@noop {} {\emph {\bibinfo {title} {Polarons}}}\ (\bibinfo
  {publisher} {Izd. Nauka},\ \bibinfo {address} {Moscow},\ \bibinfo {year}
  {1975})\BibitemShut {NoStop}%
\bibitem [{\citenamefont {Shluger}\ and\ \citenamefont
  {Stoneham}(1993)}]{SS93}%
  \BibitemOpen
  \bibfield  {author} {\bibinfo {author} {\bibfnamefont {A.~L.}\ \bibnamefont
  {Shluger}}\ and\ \bibinfo {author} {\bibfnamefont {A.~M.}\ \bibnamefont
  {Stoneham}},\ }\href@noop {} {\bibfield  {journal} {\bibinfo  {journal} {J.
  Phys. Condens. Matter}\ }\textbf {\bibinfo {volume} {5}},\ \bibinfo {pages}
  {3049} (\bibinfo {year} {1993})}\BibitemShut {NoStop}%
\bibitem [{\citenamefont {Firsov}(1995)}]{Fi95}%
  \BibitemOpen
  \bibfield  {author} {\bibinfo {author} {\bibfnamefont {Y.~A.}\ \bibnamefont
  {Firsov}},\ }\href@noop {} {\bibfield  {journal} {\bibinfo  {journal}
  {Semiconductors}\ }\textbf {\bibinfo {volume} {29}},\ \bibinfo {pages} {515}
  (\bibinfo {year} {1995})}\BibitemShut {NoStop}%
\bibitem [{\citenamefont {Fehske}\ and\ \citenamefont {Trugman}(2007)}]{FT07}%
  \BibitemOpen
  \bibfield  {author} {\bibinfo {author} {\bibfnamefont {H.}~\bibnamefont
  {Fehske}}\ and\ \bibinfo {author} {\bibfnamefont {S.~A.}\ \bibnamefont
  {Trugman}},\ }in\ \href@noop {} {\emph {\bibinfo {booktitle} {Polarons in
  Advanced Materials}}},\ \bibinfo {series} {Springer Series in Material
  Sciences}, Vol.\ \bibinfo {volume} {103},\ \bibinfo {editor} {edited by\
  \bibinfo {editor} {\bibfnamefont {A.~S.}\ \bibnamefont {Alexandrov}}}\
  (\bibinfo  {publisher} {Canopus/Springer Publishing},\ \bibinfo {address}
  {Dordrecht},\ \bibinfo {year} {2007})\ pp.\ \bibinfo {pages}
  {393--461}\BibitemShut {NoStop}%
\bibitem [{\citenamefont {Holstein}(1959{\natexlab{a}})}]{Ho59a}%
  \BibitemOpen
  \bibfield  {author} {\bibinfo {author} {\bibfnamefont {T.}~\bibnamefont
  {Holstein}},\ }\href@noop {} {\bibfield  {journal} {\bibinfo  {journal} {Ann.
  Phys. (N.Y.)}\ }\textbf {\bibinfo {volume} {8}},\ \bibinfo {pages} {325}
  (\bibinfo {year} {1959}{\natexlab{a}})}\BibitemShut {NoStop}%
\bibitem [{\citenamefont {Holstein}(1959{\natexlab{b}})}]{Ho59b}%
  \BibitemOpen
  \bibfield  {author} {\bibinfo {author} {\bibfnamefont {T.}~\bibnamefont
  {Holstein}},\ }\href@noop {} {\bibfield  {journal} {\bibinfo  {journal} {Ann.
  Phys. (N.Y.)}\ }\textbf {\bibinfo {volume} {8}},\ \bibinfo {pages} {343}
  (\bibinfo {year} {1959}{\natexlab{b}})}\BibitemShut {NoStop}%
\bibitem [{\citenamefont {Fr{\"o}hlich}(1954)}]{Fr54}%
  \BibitemOpen
  \bibfield  {author} {\bibinfo {author} {\bibfnamefont {H.}~\bibnamefont
  {Fr{\"o}hlich}},\ }\href@noop {} {\bibfield  {journal} {\bibinfo  {journal}
  {Adv. Phys.}\ }\textbf {\bibinfo {volume} {3}},\ \bibinfo {pages} {325}
  (\bibinfo {year} {1954})}\BibitemShut {NoStop}%
\bibitem [{\citenamefont {Devreese}(2006)}]{DE06}%
  \BibitemOpen
  \bibfield  {author} {\bibinfo {author} {\bibfnamefont {J.~T.}\ \bibnamefont
  {Devreese}},\ }in\ \href@noop {} {\emph {\bibinfo {booktitle} {Fr\"ohlich
  polarons from 3D to 0D, Concenpts and recent developments}}},\ \bibinfo
  {series and number} {International School of Physics Enrico Fermi},\ \bibinfo
  {editor} {edited by\ \bibinfo {editor} {\bibfnamefont {G.}~\bibnamefont
  {Iadonisi}}, \bibinfo {editor} {\bibfnamefont {J.}~\bibnamefont {Ranninger}},
  \ and\ \bibinfo {editor} {\bibfnamefont {G.}~\bibnamefont {De~Filippis}}}\
  (\bibinfo  {publisher} {IOS Press},\ \bibinfo {address} {Amsterdam},\
  \bibinfo {year} {2006})\ pp.\ \bibinfo {pages} {27--47}\BibitemShut {NoStop}%
\bibitem [{\citenamefont {Alexandrov}\ and\ \citenamefont
  {Kornilovitch}(1999)}]{AK99}%
  \BibitemOpen
  \bibfield  {author} {\bibinfo {author} {\bibfnamefont {A.~S.}\ \bibnamefont
  {Alexandrov}}\ and\ \bibinfo {author} {\bibfnamefont {P.~E.}\ \bibnamefont
  {Kornilovitch}},\ }\href@noop {} {\bibfield  {journal} {\bibinfo  {journal}
  {Phys. Rev. Lett.}\ }\textbf {\bibinfo {volume} {82}},\ \bibinfo {pages}
  {807} (\bibinfo {year} {1999})}\BibitemShut {NoStop}%
\bibitem [{\citenamefont {Fehske}\ \emph {et~al.}(2000)\citenamefont {Fehske},
  \citenamefont {Loos},\ and\ \citenamefont {Wellein}}]{FLW00}%
  \BibitemOpen
  \bibfield  {author} {\bibinfo {author} {\bibfnamefont {H.}~\bibnamefont
  {Fehske}}, \bibinfo {author} {\bibfnamefont {J.}~\bibnamefont {Loos}}, \ and\
  \bibinfo {author} {\bibfnamefont {G.}~\bibnamefont {Wellein}},\ }\href@noop
  {} {\bibfield  {journal} {\bibinfo  {journal} {Phys. Rev. B}\ }\textbf
  {\bibinfo {volume} {61}},\ \bibinfo {pages} {8016} (\bibinfo {year}
  {2000})}\BibitemShut {NoStop}%
\bibitem [{\citenamefont {Alvermann}\ \emph {et~al.}(2007)\citenamefont
  {Alvermann}, \citenamefont {Edwards},\ and\ \citenamefont {Fehske}}]{AEF07}%
  \BibitemOpen
  \bibfield  {author} {\bibinfo {author} {\bibfnamefont {A.}~\bibnamefont
  {Alvermann}}, \bibinfo {author} {\bibfnamefont {D.~M.}\ \bibnamefont
  {Edwards}}, \ and\ \bibinfo {author} {\bibfnamefont {H.}~\bibnamefont
  {Fehske}},\ }\href@noop {} {\bibfield  {journal} {\bibinfo  {journal} {Phys.
  Rev. Lett.}\ }\textbf {\bibinfo {volume} {98}},\ \bibinfo {pages} {056602}
  (\bibinfo {year} {2007})}\BibitemShut {NoStop}%
\bibitem [{\citenamefont {Chakraborty}\ \emph {et~al.}(2016)\citenamefont
  {Chakraborty}, \citenamefont {Mohanta}, \citenamefont {Taraphder},
  \citenamefont {Min},\ and\ \citenamefont {Fehske}}]{CMTMF16}%
  \BibitemOpen
  \bibfield  {author} {\bibinfo {author} {\bibfnamefont {M.}~\bibnamefont
  {Chakraborty}}, \bibinfo {author} {\bibfnamefont {N.}~\bibnamefont
  {Mohanta}}, \bibinfo {author} {\bibfnamefont {A.}~\bibnamefont {Taraphder}},
  \bibinfo {author} {\bibfnamefont {B.~I.}\ \bibnamefont {Min}}, \ and\
  \bibinfo {author} {\bibfnamefont {H.}~\bibnamefont {Fehske}},\ }\href
  {\doibase 10.1103/PhysRevB.93.155130} {\bibfield  {journal} {\bibinfo
  {journal} {Phys. Rev. B}\ }\textbf {\bibinfo {volume} {93}},\ \bibinfo
  {pages} {155130} (\bibinfo {year} {2016})}\BibitemShut {NoStop}%
\bibitem [{\citenamefont {Rashba}(1982)}]{Ra82}%
  \BibitemOpen
  \bibfield  {author} {\bibinfo {author} {\bibfnamefont {E.~I.}\ \bibnamefont
  {Rashba}},\ }in\ \href@noop {} {\emph {\bibinfo {booktitle} {Excitons}}},\
  \bibinfo {editor} {edited by\ \bibinfo {editor} {\bibfnamefont {E.~I.}\
  \bibnamefont {Rashba}}\ and\ \bibinfo {editor} {\bibfnamefont {M.~D.}\
  \bibnamefont {Sturge}}}\ (\bibinfo  {publisher} {North-Holland},\ \bibinfo
  {address} {Amsterdam},\ \bibinfo {year} {1982})\ p.\ \bibinfo {pages}
  {543}\BibitemShut {NoStop}%
\bibitem [{\citenamefont {Perlin}\ and\ \citenamefont {Wagner}(1984)}]{PW84}%
  \BibitemOpen
  \bibinfo {editor} {\bibfnamefont {Y.~E.}\ \bibnamefont {Perlin}}\ and\
  \bibinfo {editor} {\bibfnamefont {M.}~\bibnamefont {Wagner}},\ eds.,\
  \href@noop {} {\emph {\bibinfo {title} {The Dynamical Jahn-Teller Effect in
  Localized Systems}}},\ \bibinfo {series} {Modern Problems in Condensed Matter
  Sciences}\ No.~\bibinfo {number} {7}\ (\bibinfo  {publisher}
  {North-Holland},\ \bibinfo {address} {Amsterdam},\ \bibinfo {year}
  {1984})\BibitemShut {NoStop}%
\bibitem [{\citenamefont {El~Shawish}\ \emph {et~al.}(2003)\citenamefont
  {El~Shawish}, \citenamefont {Bon\v{c}a}, \citenamefont {Ku},\ and\
  \citenamefont {Trugman}}]{EBKT03}%
  \BibitemOpen
  \bibfield  {author} {\bibinfo {author} {\bibfnamefont {S.}~\bibnamefont
  {El~Shawish}}, \bibinfo {author} {\bibfnamefont {J.}~\bibnamefont
  {Bon\v{c}a}}, \bibinfo {author} {\bibfnamefont {L.-C.}\ \bibnamefont {Ku}}, \
  and\ \bibinfo {author} {\bibfnamefont {S.~A.}\ \bibnamefont {Trugman}},\
  }\href@noop {} {\bibfield  {journal} {\bibinfo  {journal} {Phys. Rev. B}\
  }\textbf {\bibinfo {volume} {67}},\ \bibinfo {pages} {014301} (\bibinfo
  {year} {2003})}\BibitemShut {NoStop}%
\bibitem [{\citenamefont {Iadonisi}\ \emph {et~al.}(2006)\citenamefont
  {Iadonisi}, \citenamefont {Ranninger},\ and\ \citenamefont
  {Filipis}}]{IRF06}%
  \BibitemOpen
  \bibinfo {editor} {\bibfnamefont {G.}~\bibnamefont {Iadonisi}}, \bibinfo
  {editor} {\bibfnamefont {J.}~\bibnamefont {Ranninger}}, \ and\ \bibinfo
  {editor} {\bibfnamefont {G.~D.}\ \bibnamefont {Filipis}},\ eds.,\ \href@noop
  {} {\emph {\bibinfo {title} {Polarons in bulk materials and systems with
  reduced dimensionality}}},\ \bibinfo {series} {International School of
  Physics Enrico Fermi}, Vol.\ \bibinfo {volume} {161}\ (\bibinfo  {publisher}
  {IOS Press},\ \bibinfo {address} {Amsterdam},\ \bibinfo {year}
  {2006})\BibitemShut {NoStop}%
\bibitem [{\citenamefont {Egami}(2006)}]{Eg06}%
  \BibitemOpen
  \bibfield  {author} {\bibinfo {author} {\bibfnamefont {T.}~\bibnamefont
  {Egami}},\ }in\ \href@noop {} {\emph {\bibinfo {booktitle} {Polarons in bulk
  materials and systems with reduced dimensionality}}},\ \bibinfo {series}
  {International School of Physics Enrico Fermi}, Vol.\ \bibinfo {volume}
  {161},\ \bibinfo {editor} {edited by\ \bibinfo {editor} {\bibfnamefont
  {G.}~\bibnamefont {Iadonisi}}, \bibinfo {editor} {\bibfnamefont
  {J.}~\bibnamefont {Ranninger}}, \ and\ \bibinfo {editor} {\bibfnamefont
  {G.~D.}\ \bibnamefont {Filipis}}}\ (\bibinfo  {publisher} {IOS Press},\
  \bibinfo {address} {Amsterdam},\ \bibinfo {year} {2006})\ pp.\ \bibinfo
  {pages} {101--117}\BibitemShut {NoStop}%
\bibitem [{\citenamefont {Salje}\ \emph {et~al.}(1995)\citenamefont {Salje},
  \citenamefont {Alexandrov},\ and\ \citenamefont {Liang}}]{SAL95}%
  \BibitemOpen
  \bibfield  {author} {\bibinfo {author} {\bibfnamefont {E.~K.~H.}\
  \bibnamefont {Salje}}, \bibinfo {author} {\bibfnamefont {A.~S.}\ \bibnamefont
  {Alexandrov}}, \ and\ \bibinfo {author} {\bibfnamefont {W.~Y.}\ \bibnamefont
  {Liang}},\ }\href@noop {} {\emph {\bibinfo {title} {Polarons and Bipolarons
  in High Temperature Superconductors and Related Materials}}}\ (\bibinfo
  {publisher} {Cambridge University Press},\ \bibinfo {address} {Cambridge},\
  \bibinfo {year} {1995})\BibitemShut {NoStop}%
\bibitem [{\citenamefont {Calvani}\ \emph {et~al.}(1997)\citenamefont
  {Calvani}, \citenamefont {Dore}, \citenamefont {Lupi}, \citenamefont
  {Paolone}, \citenamefont {Maselli}, \citenamefont {Guira}, \citenamefont
  {Ruzicka}, \citenamefont {Cheong},\ and\ \citenamefont {Sadowski}}]{Caea97}%
  \BibitemOpen
  \bibfield  {author} {\bibinfo {author} {\bibfnamefont {P.}~\bibnamefont
  {Calvani}}, \bibinfo {author} {\bibfnamefont {P.}~\bibnamefont {Dore}},
  \bibinfo {author} {\bibfnamefont {S.}~\bibnamefont {Lupi}}, \bibinfo {author}
  {\bibfnamefont {A.}~\bibnamefont {Paolone}}, \bibinfo {author} {\bibfnamefont
  {P.}~\bibnamefont {Maselli}}, \bibinfo {author} {\bibfnamefont
  {P.}~\bibnamefont {Guira}}, \bibinfo {author} {\bibfnamefont
  {B.}~\bibnamefont {Ruzicka}}, \bibinfo {author} {\bibfnamefont {S.-W.}\
  \bibnamefont {Cheong}}, \ and\ \bibinfo {author} {\bibfnamefont
  {W.}~\bibnamefont {Sadowski}},\ }\href@noop {} {\bibfield  {journal}
  {\bibinfo  {journal} {J. Supercond.}\ }\textbf {\bibinfo {volume} {10}},\
  \bibinfo {pages} {293} (\bibinfo {year} {1997})}\BibitemShut {NoStop}%
\bibitem [{\citenamefont {Alexandrov}(2007)}]{ALex07}%
  \BibitemOpen
  \bibinfo {editor} {\bibfnamefont {A.~S.}\ \bibnamefont {Alexandrov}},\ ed.,\
  \href@noop {} {\emph {\bibinfo {title} {Polarons in Advanced Materials}}},\
  \bibinfo {series} {Springer Series in Material Sciences}, Vol.\ \bibinfo
  {volume} {103}\ (\bibinfo  {publisher} {Springer},\ \bibinfo {address}
  {Dordrecht},\ \bibinfo {year} {2007})\BibitemShut {NoStop}%
\bibitem [{\citenamefont {Bar-Yam}\ \emph {et~al.}(1992)\citenamefont
  {Bar-Yam}, \citenamefont {Egami}, \citenamefont {de~Leon},\ and\
  \citenamefont {Bishop}}]{BEMB92}%
  \BibitemOpen
  \bibfield  {author} {\bibinfo {author} {\bibfnamefont {Y.}~\bibnamefont
  {Bar-Yam}}, \bibinfo {author} {\bibfnamefont {T.}~\bibnamefont {Egami}},
  \bibinfo {author} {\bibfnamefont {J.~M.}\ \bibnamefont {de~Leon}}, \ and\
  \bibinfo {author} {\bibfnamefont {A.~R.}\ \bibnamefont {Bishop}},\
  }\href@noop {} {\emph {\bibinfo {title} {Lattice Effects in High--$T_c$
  Superconductors}}}\ (\bibinfo  {publisher} {World Scientific},\ \bibinfo
  {address} {Singapore},\ \bibinfo {year} {1992})\BibitemShut {NoStop}%
\bibitem [{\citenamefont {Mott}(1993)}]{Mott93}%
  \BibitemOpen
  \bibfield  {author} {\bibinfo {author} {\bibfnamefont {N.~F.}\ \bibnamefont
  {Mott}},\ }\href {\doibase 10.1088/0953-8984/5/22/003} {\bibfield  {journal}
  {\bibinfo  {journal} {Journal of Physics: Condensed Matter}\ }\textbf
  {\bibinfo {volume} {5}},\ \bibinfo {pages} {3487} (\bibinfo {year}
  {1993})}\BibitemShut {NoStop}%
\bibitem [{\citenamefont {Bi}\ and\ \citenamefont {Eklund}(1993)}]{BE93}%
  \BibitemOpen
  \bibfield  {author} {\bibinfo {author} {\bibfnamefont {X.-X.}\ \bibnamefont
  {Bi}}\ and\ \bibinfo {author} {\bibfnamefont {P.~C.}\ \bibnamefont
  {Eklund}},\ }\href@noop {} {\bibfield  {journal} {\bibinfo  {journal} {Phys.
  Rev. Lett.}\ }\textbf {\bibinfo {volume} {70}},\ \bibinfo {pages} {2625}
  (\bibinfo {year} {1993})}\BibitemShut {NoStop}%
\bibitem [{\citenamefont {Tyunina}\ \emph {et~al.}(2023)\citenamefont
  {Tyunina}, \citenamefont {Savinov}, \citenamefont {Pacherova},\ and\
  \citenamefont {Dejneka}}]{TSPD23}%
  \BibitemOpen
  \bibfield  {author} {\bibinfo {author} {\bibfnamefont {M.}~\bibnamefont
  {Tyunina}}, \bibinfo {author} {\bibfnamefont {M.}~\bibnamefont {Savinov}},
  \bibinfo {author} {\bibfnamefont {O.}~\bibnamefont {Pacherova}}, \ and\
  \bibinfo {author} {\bibfnamefont {A.}~\bibnamefont {Dejneka}},\ }\href@noop
  {} {\bibfield  {journal} {\bibinfo  {journal} {Scientitic Report}\ }\textbf
  {\bibinfo {volume} {13}},\ \bibinfo {pages} {12493} (\bibinfo {year}
  {2023})}\BibitemShut {NoStop}%
\bibitem [{\citenamefont {Millis}\ \emph {et~al.}(1995)\citenamefont {Millis},
  \citenamefont {Littlewood},\ and\ \citenamefont {Shraiman}}]{MLS95}%
  \BibitemOpen
  \bibfield  {author} {\bibinfo {author} {\bibfnamefont {A.~J.}\ \bibnamefont
  {Millis}}, \bibinfo {author} {\bibfnamefont {P.~B.}\ \bibnamefont
  {Littlewood}}, \ and\ \bibinfo {author} {\bibfnamefont {B.~I.}\ \bibnamefont
  {Shraiman}},\ }\href@noop {} {\bibfield  {journal} {\bibinfo  {journal}
  {Phys. Rev. Lett.}\ }\textbf {\bibinfo {volume} {74}},\ \bibinfo {pages}
  {5144} (\bibinfo {year} {1995})}\BibitemShut {NoStop}%
\bibitem [{\citenamefont {Jaime}\ \emph {et~al.}(1997)\citenamefont {Jaime},
  \citenamefont {Hardner}, \citenamefont {Salamon}, \citenamefont {Rubinstein},
  \citenamefont {Dorsey},\ and\ \citenamefont {Emin}}]{JHSRDE97}%
  \BibitemOpen
  \bibfield  {author} {\bibinfo {author} {\bibfnamefont {M.}~\bibnamefont
  {Jaime}}, \bibinfo {author} {\bibfnamefont {H.~T.}\ \bibnamefont {Hardner}},
  \bibinfo {author} {\bibfnamefont {M.~B.}\ \bibnamefont {Salamon}}, \bibinfo
  {author} {\bibfnamefont {M.}~\bibnamefont {Rubinstein}}, \bibinfo {author}
  {\bibfnamefont {P.}~\bibnamefont {Dorsey}}, \ and\ \bibinfo {author}
  {\bibfnamefont {D.}~\bibnamefont {Emin}},\ }\href@noop {} {\bibfield
  {journal} {\bibinfo  {journal} {Phys. Rev. Lett.}\ }\textbf {\bibinfo
  {volume} {78}},\ \bibinfo {pages} {951} (\bibinfo {year} {1997})}\BibitemShut
  {NoStop}%
\bibitem [{\citenamefont {Louca}\ and\ \citenamefont {Egami}(1999)}]{LE99}%
  \BibitemOpen
  \bibfield  {author} {\bibinfo {author} {\bibfnamefont {D.}~\bibnamefont
  {Louca}}\ and\ \bibinfo {author} {\bibfnamefont {T.}~\bibnamefont {Egami}},\
  }\href@noop {} {\bibfield  {journal} {\bibinfo  {journal} {Phys. Rev. B}\
  }\textbf {\bibinfo {volume} {59}},\ \bibinfo {pages} {6193} (\bibinfo {year}
  {1999})}\BibitemShut {NoStop}%
\bibitem [{\citenamefont {Gerlach}\ and\ \citenamefont {L\"owen}(1991)}]{GL91}%
  \BibitemOpen
  \bibfield  {author} {\bibinfo {author} {\bibfnamefont {B.}~\bibnamefont
  {Gerlach}}\ and\ \bibinfo {author} {\bibfnamefont {H.}~\bibnamefont
  {L\"owen}},\ }\href@noop {} {\bibfield  {journal} {\bibinfo  {journal} {Rev.
  Mod. Phys.}\ }\textbf {\bibinfo {volume} {63}},\ \bibinfo {pages} {63}
  (\bibinfo {year} {1991})}\BibitemShut {NoStop}%
\bibitem [{\citenamefont {Wellein}\ and\ \citenamefont {Fehske}(1998)}]{WF98a}%
  \BibitemOpen
  \bibfield  {author} {\bibinfo {author} {\bibfnamefont {G.}~\bibnamefont
  {Wellein}}\ and\ \bibinfo {author} {\bibfnamefont {H.}~\bibnamefont
  {Fehske}},\ }\href@noop {} {\bibfield  {journal} {\bibinfo  {journal} {Phys.
  Rev. B}\ }\textbf {\bibinfo {volume} {58}},\ \bibinfo {pages} {6208}
  (\bibinfo {year} {1998})}\BibitemShut {NoStop}%
\bibitem [{\citenamefont {De~Raedt}\ and\ \citenamefont
  {Lagendijk}(1982)}]{RL82}%
  \BibitemOpen
  \bibfield  {author} {\bibinfo {author} {\bibfnamefont {H.}~\bibnamefont
  {De~Raedt}}\ and\ \bibinfo {author} {\bibfnamefont {A.}~\bibnamefont
  {Lagendijk}},\ }\href@noop {} {\bibfield  {journal} {\bibinfo  {journal}
  {Phys. Rev. Lett.}\ }\textbf {\bibinfo {volume} {49}},\ \bibinfo {pages}
  {1522} (\bibinfo {year} {1982})}\BibitemShut {NoStop}%
\bibitem [{\citenamefont {Berger}\ \emph {et~al.}(1995)\citenamefont {Berger},
  \citenamefont {Val\'{a}\v{s}ek},\ and\ \citenamefont {v.~d. Linden}}]{BVL95}%
  \BibitemOpen
  \bibfield  {author} {\bibinfo {author} {\bibfnamefont {E.}~\bibnamefont
  {Berger}}, \bibinfo {author} {\bibfnamefont {P.}~\bibnamefont
  {Val\'{a}\v{s}ek}}, \ and\ \bibinfo {author} {\bibfnamefont {W.}~\bibnamefont
  {v.~d. Linden}},\ }\href@noop {} {\bibfield  {journal} {\bibinfo  {journal}
  {Phys. Rev. B}\ }\textbf {\bibinfo {volume} {52}},\ \bibinfo {pages} {4806}
  (\bibinfo {year} {1995})}\BibitemShut {NoStop}%
\bibitem [{\citenamefont {Kornilovitch}\ and\ \citenamefont
  {Pike}(1997)}]{KP97}%
  \BibitemOpen
  \bibfield  {author} {\bibinfo {author} {\bibfnamefont {P.~E.}\ \bibnamefont
  {Kornilovitch}}\ and\ \bibinfo {author} {\bibfnamefont {E.~R.}\ \bibnamefont
  {Pike}},\ }\href@noop {} {\bibfield  {journal} {\bibinfo  {journal} {Phys.
  Rev. B}\ }\textbf {\bibinfo {volume} {55}},\ \bibinfo {pages} {R8634}
  (\bibinfo {year} {1997})}\BibitemShut {NoStop}%
\bibitem [{\citenamefont {Mishchenko}\ \emph {et~al.}(2000)\citenamefont
  {Mishchenko}, \citenamefont {Prokof'ev}, \citenamefont {Sakamoto},\ and\
  \citenamefont {Svistunov}}]{MPSS00}%
  \BibitemOpen
  \bibfield  {author} {\bibinfo {author} {\bibfnamefont {A.~S.}\ \bibnamefont
  {Mishchenko}}, \bibinfo {author} {\bibfnamefont {N.~V.}\ \bibnamefont
  {Prokof'ev}}, \bibinfo {author} {\bibfnamefont {A.}~\bibnamefont {Sakamoto}},
  \ and\ \bibinfo {author} {\bibfnamefont {B.~V.}\ \bibnamefont {Svistunov}},\
  }\href@noop {} {\bibfield  {journal} {\bibinfo  {journal} {Phys. Rev. B}\
  }\textbf {\bibinfo {volume} {62}},\ \bibinfo {pages} {6317} (\bibinfo {year}
  {2000})}\BibitemShut {NoStop}%
\bibitem [{\citenamefont {Hohenadler}\ \emph {et~al.}(2004)\citenamefont
  {Hohenadler}, \citenamefont {Evertz},\ and\ \citenamefont {von~der
  Linden}}]{HEL04}%
  \BibitemOpen
  \bibfield  {author} {\bibinfo {author} {\bibfnamefont {M.}~\bibnamefont
  {Hohenadler}}, \bibinfo {author} {\bibfnamefont {H.~G.}\ \bibnamefont
  {Evertz}}, \ and\ \bibinfo {author} {\bibfnamefont {W.}~\bibnamefont {von~der
  Linden}},\ }\href@noop {} {\bibfield  {journal} {\bibinfo  {journal} {Phys.
  Rev. B}\ }\textbf {\bibinfo {volume} {69}},\ \bibinfo {pages} {024301}
  (\bibinfo {year} {2004})}\BibitemShut {NoStop}%
\bibitem [{\citenamefont {Burovski}\ \emph {et~al.}(2008)\citenamefont
  {Burovski}, \citenamefont {Fehske},\ and\ \citenamefont
  {Mishchenko}}]{BFM08}%
  \BibitemOpen
  \bibfield  {author} {\bibinfo {author} {\bibfnamefont {E.}~\bibnamefont
  {Burovski}}, \bibinfo {author} {\bibfnamefont {H.}~\bibnamefont {Fehske}}, \
  and\ \bibinfo {author} {\bibfnamefont {A.~S.}\ \bibnamefont {Mishchenko}},\
  }\href@noop {} {\bibfield  {journal} {\bibinfo  {journal} {Phys. Rev. Lett.}\
  }\textbf {\bibinfo {volume} {101}},\ \bibinfo {pages} {116403} (\bibinfo
  {year} {2008})}\BibitemShut {NoStop}%
\bibitem [{\citenamefont {Mishchenko}\ \emph {et~al.}(2009)\citenamefont
  {Mishchenko}, \citenamefont {Nagaosa}, \citenamefont {Alvermann},
  \citenamefont {Fehske}, \citenamefont {De~Filippis}, \citenamefont
  {Cataudella},\ and\ \citenamefont {Sushkov}}]{MNAFDCS09}%
  \BibitemOpen
  \bibfield  {author} {\bibinfo {author} {\bibfnamefont {A.~S.}\ \bibnamefont
  {Mishchenko}}, \bibinfo {author} {\bibfnamefont {N.}~\bibnamefont {Nagaosa}},
  \bibinfo {author} {\bibfnamefont {A.}~\bibnamefont {Alvermann}}, \bibinfo
  {author} {\bibfnamefont {H.}~\bibnamefont {Fehske}}, \bibinfo {author}
  {\bibfnamefont {G.}~\bibnamefont {De~Filippis}}, \bibinfo {author}
  {\bibfnamefont {V.}~\bibnamefont {Cataudella}}, \ and\ \bibinfo {author}
  {\bibfnamefont {O.~P.}\ \bibnamefont {Sushkov}},\ }\href {\doibase
  10.1103/PhysRevB.79.180301} {\bibfield  {journal} {\bibinfo  {journal} {Phys.
  Rev. B}\ }\textbf {\bibinfo {volume} {79}},\ \bibinfo {pages} {180301}
  (\bibinfo {year} {2009})}\BibitemShut {NoStop}%
\bibitem [{\citenamefont {Hohenadler}(2016)}]{Ho16}%
  \BibitemOpen
  \bibfield  {author} {\bibinfo {author} {\bibfnamefont {M.}~\bibnamefont
  {Hohenadler}},\ }\href {\doibase 10.1103/PhysRevLett.117.206404} {\bibfield
  {journal} {\bibinfo  {journal} {Phys. Rev. Lett.}\ }\textbf {\bibinfo
  {volume} {117}},\ \bibinfo {pages} {206404} (\bibinfo {year}
  {2016})}\BibitemShut {NoStop}%
\bibitem [{\citenamefont {Wellein}\ \emph {et~al.}(1996)\citenamefont
  {Wellein}, \citenamefont {R\"oder},\ and\ \citenamefont {Fehske}}]{WRF96}%
  \BibitemOpen
  \bibfield  {author} {\bibinfo {author} {\bibfnamefont {G.}~\bibnamefont
  {Wellein}}, \bibinfo {author} {\bibfnamefont {H.}~\bibnamefont {R\"oder}}, \
  and\ \bibinfo {author} {\bibfnamefont {H.}~\bibnamefont {Fehske}},\
  }\href@noop {} {\bibfield  {journal} {\bibinfo  {journal} {Phys. Rev. B}\
  }\textbf {\bibinfo {volume} {53}},\ \bibinfo {pages} {9666} (\bibinfo {year}
  {1996})}\BibitemShut {NoStop}%
\bibitem [{\citenamefont {Stephan}(1996)}]{St96}%
  \BibitemOpen
  \bibfield  {author} {\bibinfo {author} {\bibfnamefont {W.}~\bibnamefont
  {Stephan}},\ }\href@noop {} {\bibfield  {journal} {\bibinfo  {journal} {Phys.
  Rev. B}\ }\textbf {\bibinfo {volume} {54}},\ \bibinfo {pages} {8981}
  (\bibinfo {year} {1996})}\BibitemShut {NoStop}%
\bibitem [{\citenamefont {Bon\v{c}a}\ \emph {et~al.}(1999)\citenamefont
  {Bon\v{c}a}, \citenamefont {Trugman},\ and\ \citenamefont
  {Batisti\'{c}}}]{BTB99}%
  \BibitemOpen
  \bibfield  {author} {\bibinfo {author} {\bibfnamefont {J.}~\bibnamefont
  {Bon\v{c}a}}, \bibinfo {author} {\bibfnamefont {S.~A.}\ \bibnamefont
  {Trugman}}, \ and\ \bibinfo {author} {\bibfnamefont {I.}~\bibnamefont
  {Batisti\'{c}}},\ }\href@noop {} {\bibfield  {journal} {\bibinfo  {journal}
  {Phys. Rev. B}\ }\textbf {\bibinfo {volume} {60}},\ \bibinfo {pages} {1633}
  (\bibinfo {year} {1999})}\BibitemShut {NoStop}%
\bibitem [{\citenamefont {Ku}\ \emph {et~al.}(2002)\citenamefont {Ku},
  \citenamefont {Trugman},\ and\ \citenamefont {Bon\v{c}a}}]{KTB02}%
  \BibitemOpen
  \bibfield  {author} {\bibinfo {author} {\bibfnamefont {L.-C.}\ \bibnamefont
  {Ku}}, \bibinfo {author} {\bibfnamefont {S.~A.}\ \bibnamefont {Trugman}}, \
  and\ \bibinfo {author} {\bibfnamefont {J.}~\bibnamefont {Bon\v{c}a}},\
  }\href@noop {} {\bibfield  {journal} {\bibinfo  {journal} {Phys. Rev. B}\
  }\textbf {\bibinfo {volume} {65}},\ \bibinfo {pages} {174306} (\bibinfo
  {year} {2002})}\BibitemShut {NoStop}%
\bibitem [{\citenamefont {Alvermann}\ \emph {et~al.}(2008)\citenamefont
  {Alvermann}, \citenamefont {Fehske},\ and\ \citenamefont {Trugman}}]{AFT08}%
  \BibitemOpen
  \bibfield  {author} {\bibinfo {author} {\bibfnamefont {A.}~\bibnamefont
  {Alvermann}}, \bibinfo {author} {\bibfnamefont {H.}~\bibnamefont {Fehske}}, \
  and\ \bibinfo {author} {\bibfnamefont {S.~A.}\ \bibnamefont {Trugman}},\
  }\href@noop {} {\bibfield  {journal} {\bibinfo  {journal} {Phys. Rev. B}\
  }\textbf {\bibinfo {volume} {78}},\ \bibinfo {pages} {165106} (\bibinfo
  {year} {2008})}\BibitemShut {NoStop}%
\bibitem [{\citenamefont {Alvermann}\ \emph {et~al.}(2010)\citenamefont
  {Alvermann}, \citenamefont {Fehske},\ and\ \citenamefont {Trugman}}]{AFT10}%
  \BibitemOpen
  \bibfield  {author} {\bibinfo {author} {\bibfnamefont {A.}~\bibnamefont
  {Alvermann}}, \bibinfo {author} {\bibfnamefont {H.}~\bibnamefont {Fehske}}, \
  and\ \bibinfo {author} {\bibfnamefont {S.~A.}\ \bibnamefont {Trugman}},\
  }\href@noop {} {\bibfield  {journal} {\bibinfo  {journal} {Phys. Rev. B}\
  }\textbf {\bibinfo {volume} {81}},\ \bibinfo {pages} {165113} (\bibinfo
  {year} {2010})}\BibitemShut {NoStop}%
\bibitem [{\citenamefont {Li}\ \emph {et~al.}(2010)\citenamefont {Li},
  \citenamefont {Baillie}, \citenamefont {Blois},\ and\ \citenamefont
  {Marsiglio}}]{LBBM10}%
  \BibitemOpen
  \bibfield  {author} {\bibinfo {author} {\bibfnamefont {Z.}~\bibnamefont
  {Li}}, \bibinfo {author} {\bibfnamefont {D.}~\bibnamefont {Baillie}},
  \bibinfo {author} {\bibfnamefont {C.}~\bibnamefont {Blois}}, \ and\ \bibinfo
  {author} {\bibfnamefont {F.}~\bibnamefont {Marsiglio}},\ }\href {\doibase
  10.1103/PhysRevB.81.115114} {\bibfield  {journal} {\bibinfo  {journal} {Phys.
  Rev. B}\ }\textbf {\bibinfo {volume} {81}},\ \bibinfo {pages} {115114}
  (\bibinfo {year} {2010})}\BibitemShut {NoStop}%
\bibitem [{\citenamefont {Chakraborty}\ \emph {et~al.}(2012)\citenamefont
  {Chakraborty}, \citenamefont {Min}, \citenamefont {Chakrabarti},\ and\
  \citenamefont {Das}}]{CMCD12}%
  \BibitemOpen
  \bibfield  {author} {\bibinfo {author} {\bibfnamefont {M.}~\bibnamefont
  {Chakraborty}}, \bibinfo {author} {\bibfnamefont {B.~I.}\ \bibnamefont
  {Min}}, \bibinfo {author} {\bibfnamefont {A.}~\bibnamefont {Chakrabarti}}, \
  and\ \bibinfo {author} {\bibfnamefont {A.~N.}\ \bibnamefont {Das}},\ }\href
  {\doibase 10.1103/PhysRevB.85.245127} {\bibfield  {journal} {\bibinfo
  {journal} {Phys. Rev. B}\ }\textbf {\bibinfo {volume} {85}},\ \bibinfo
  {pages} {245127} (\bibinfo {year} {2012})}\BibitemShut {NoStop}%
\bibitem [{\citenamefont {Chakraborty}\ \emph {et~al.}(2014)\citenamefont
  {Chakraborty}, \citenamefont {Tezuka},\ and\ \citenamefont {Min}}]{CTM14}%
  \BibitemOpen
  \bibfield  {author} {\bibinfo {author} {\bibfnamefont {M.}~\bibnamefont
  {Chakraborty}}, \bibinfo {author} {\bibfnamefont {M.}~\bibnamefont {Tezuka}},
  \ and\ \bibinfo {author} {\bibfnamefont {B.~I.}\ \bibnamefont {Min}},\ }\href
  {\doibase 10.1103/PhysRevB.89.035146} {\bibfield  {journal} {\bibinfo
  {journal} {Phys. Rev. B}\ }\textbf {\bibinfo {volume} {89}},\ \bibinfo
  {pages} {035146} (\bibinfo {year} {2014})}\BibitemShut {NoStop}%
\bibitem [{\citenamefont {Zhang}\ \emph {et~al.}(1998)\citenamefont {Zhang},
  \citenamefont {Jeckelmann},\ and\ \citenamefont {White}}]{ZJW98}%
  \BibitemOpen
  \bibfield  {author} {\bibinfo {author} {\bibfnamefont {C.}~\bibnamefont
  {Zhang}}, \bibinfo {author} {\bibfnamefont {E.}~\bibnamefont {Jeckelmann}}, \
  and\ \bibinfo {author} {\bibfnamefont {S.~R.}\ \bibnamefont {White}},\
  }\href@noop {} {\bibfield  {journal} {\bibinfo  {journal} {Phys. Rev. Lett.}\
  }\textbf {\bibinfo {volume} {80}},\ \bibinfo {pages} {2661} (\bibinfo {year}
  {1998})}\BibitemShut {NoStop}%
\bibitem [{\citenamefont {Jeckelmann}\ and\ \citenamefont
  {White}(1998{\natexlab{b}})}]{JW98b}%
  \BibitemOpen
  \bibfield  {author} {\bibinfo {author} {\bibfnamefont {E.}~\bibnamefont
  {Jeckelmann}}\ and\ \bibinfo {author} {\bibfnamefont {S.~R.}\ \bibnamefont
  {White}},\ }\href@noop {} {\bibfield  {journal} {\bibinfo  {journal} {Phys.
  Rev. B}\ }\textbf {\bibinfo {volume} {57}},\ \bibinfo {pages} {6376}
  (\bibinfo {year} {1998}{\natexlab{b}})}\BibitemShut {NoStop}%
\bibitem [{\citenamefont {Jeckelmann}\ and\ \citenamefont
  {Fehske}(2007)}]{JF07}%
  \BibitemOpen
  \bibfield  {author} {\bibinfo {author} {\bibfnamefont {E.}~\bibnamefont
  {Jeckelmann}}\ and\ \bibinfo {author} {\bibfnamefont {H.}~\bibnamefont
  {Fehske}},\ }\href@noop {} {\bibfield  {journal} {\bibinfo  {journal}
  {Rivista del Nuovo Cimento}\ }\textbf {\bibinfo {volume} {30}},\ \bibinfo
  {pages} {259} (\bibinfo {year} {2007})}\BibitemShut {NoStop}%
\bibitem [{\citenamefont {Wei{\ss}e}\ \emph {et~al.}(2006)\citenamefont
  {Wei{\ss}e}, \citenamefont {Wellein}, \citenamefont {Alvermann},\ and\
  \citenamefont {Fehske}}]{wwaf06}%
  \BibitemOpen
  \bibfield  {author} {\bibinfo {author} {\bibfnamefont {A.}~\bibnamefont
  {Wei{\ss}e}}, \bibinfo {author} {\bibfnamefont {G.}~\bibnamefont {Wellein}},
  \bibinfo {author} {\bibfnamefont {A.}~\bibnamefont {Alvermann}}, \ and\
  \bibinfo {author} {\bibfnamefont {H.}~\bibnamefont {Fehske}},\ }\href
  {\doibase 10.1103/RevModPhys.78.275} {\bibfield  {journal} {\bibinfo
  {journal} {Rev. Mod. Phys.}\ }\textbf {\bibinfo {volume} {78}},\ \bibinfo
  {pages} {275} (\bibinfo {year} {2006})}\BibitemShut {NoStop}%
\bibitem [{\citenamefont {Su}\ \emph {et~al.}(1979)\citenamefont {Su},
  \citenamefont {Schrieffer},\ and\ \citenamefont {Heeger}}]{SSH79}%
  \BibitemOpen
  \bibfield  {author} {\bibinfo {author} {\bibfnamefont {W.~P.}\ \bibnamefont
  {Su}}, \bibinfo {author} {\bibfnamefont {J.~R.}\ \bibnamefont {Schrieffer}},
  \ and\ \bibinfo {author} {\bibfnamefont {A.~J.}\ \bibnamefont {Heeger}},\
  }\href@noop {} {\bibfield  {journal} {\bibinfo  {journal} {Phys. Rev. Lett.}\
  }\textbf {\bibinfo {volume} {42}},\ \bibinfo {pages} {1698} (\bibinfo {year}
  {1979})}\BibitemShut {NoStop}%
\bibitem [{\citenamefont {Marchand}\ \emph {et~al.}(2010)\citenamefont
  {Marchand}, \citenamefont {De~Filippis}, \citenamefont {Cataudella},
  \citenamefont {Berciu}, \citenamefont {Nagaosa}, \citenamefont {Prokof'ev},
  \citenamefont {Mishchenko},\ and\ \citenamefont {Stamp}}]{MDCBNPMS10}%
  \BibitemOpen
  \bibfield  {author} {\bibinfo {author} {\bibfnamefont {D.~J.~J.}\
  \bibnamefont {Marchand}}, \bibinfo {author} {\bibfnamefont {G.}~\bibnamefont
  {De~Filippis}}, \bibinfo {author} {\bibfnamefont {V.}~\bibnamefont
  {Cataudella}}, \bibinfo {author} {\bibfnamefont {M.}~\bibnamefont {Berciu}},
  \bibinfo {author} {\bibfnamefont {N.}~\bibnamefont {Nagaosa}}, \bibinfo
  {author} {\bibfnamefont {N.~V.}\ \bibnamefont {Prokof'ev}}, \bibinfo {author}
  {\bibfnamefont {A.~S.}\ \bibnamefont {Mishchenko}}, \ and\ \bibinfo {author}
  {\bibfnamefont {P.~C.~E.}\ \bibnamefont {Stamp}},\ }\href@noop {} {\bibfield
  {journal} {\bibinfo  {journal} {Phys. Rev. Lett.}\ }\textbf {\bibinfo
  {volume} {105}},\ \bibinfo {pages} {266605} (\bibinfo {year}
  {2010})}\BibitemShut {NoStop}%
\bibitem [{\citenamefont {Chakraborty}\ \emph {et~al.}(2011)\citenamefont
  {Chakraborty}, \citenamefont {Das},\ and\ \citenamefont
  {Chakrabarti}}]{CDC11}%
  \BibitemOpen
  \bibfield  {author} {\bibinfo {author} {\bibfnamefont {M.}~\bibnamefont
  {Chakraborty}}, \bibinfo {author} {\bibfnamefont {A.~N.}\ \bibnamefont
  {Das}}, \ and\ \bibinfo {author} {\bibfnamefont {A.}~\bibnamefont
  {Chakrabarti}},\ }\href@noop {} {\bibfield  {journal} {\bibinfo  {journal}
  {J. Phys. Condens. Matter}\ }\textbf {\bibinfo {volume} {23}},\ \bibinfo
  {pages} {025601} (\bibinfo {year} {2011})}\BibitemShut {NoStop}%
\bibitem [{\citenamefont {Chandler}\ \emph {et~al.}(2016)\citenamefont
  {Chandler}, \citenamefont {Prosko},\ and\ \citenamefont {Marsiglio}}]{CPM16}%
  \BibitemOpen
  \bibfield  {author} {\bibinfo {author} {\bibfnamefont {C.~J.}\ \bibnamefont
  {Chandler}}, \bibinfo {author} {\bibfnamefont {C.}~\bibnamefont {Prosko}}, \
  and\ \bibinfo {author} {\bibfnamefont {F.}~\bibnamefont {Marsiglio}},\ }\href
  {https://doi.org/10.1038/srep32591} {\bibfield  {journal} {\bibinfo
  {journal} {Sci. Rep.}\ }\textbf {\bibinfo {volume} {6}},\ \bibinfo {pages}
  {32591} (\bibinfo {year} {2016})}\BibitemShut {NoStop}%
\bibitem [{\citenamefont {Wang}\ \emph {et~al.}(2021)\citenamefont {Wang},
  \citenamefont {Chen}, \citenamefont {Shi}, \citenamefont {Moritz},
  \citenamefont {Shen},\ and\ \citenamefont {Devereaux}}]{WCSMSD21}%
  \BibitemOpen
  \bibfield  {author} {\bibinfo {author} {\bibfnamefont {Y.}~\bibnamefont
  {Wang}}, \bibinfo {author} {\bibfnamefont {Z.}~\bibnamefont {Chen}}, \bibinfo
  {author} {\bibfnamefont {T.}~\bibnamefont {Shi}}, \bibinfo {author}
  {\bibfnamefont {B.}~\bibnamefont {Moritz}}, \bibinfo {author} {\bibfnamefont
  {Z.-X.}\ \bibnamefont {Shen}}, \ and\ \bibinfo {author} {\bibfnamefont
  {T.~P.}\ \bibnamefont {Devereaux}},\ }\href {\doibase
  10.1103/PhysRevLett.127.197003} {\bibfield  {journal} {\bibinfo  {journal}
  {Phys. Rev. Lett.}\ }\textbf {\bibinfo {volume} {127}},\ \bibinfo {pages}
  {197003} (\bibinfo {year} {2021})}\BibitemShut {NoStop}%
\bibitem [{\citenamefont {Tang}\ \emph {et~al.}(2023)\citenamefont {Tang},
  \citenamefont {Moritz}, \citenamefont {Peng}, \citenamefont {Shen},\ and\
  \citenamefont {Devereaux}}]{TMPSD21}%
  \BibitemOpen
  \bibfield  {author} {\bibinfo {author} {\bibfnamefont {T.}~\bibnamefont
  {Tang}}, \bibinfo {author} {\bibfnamefont {B.}~\bibnamefont {Moritz}},
  \bibinfo {author} {\bibfnamefont {C.}~\bibnamefont {Peng}}, \bibinfo {author}
  {\bibfnamefont {Z.-X.}\ \bibnamefont {Shen}}, \ and\ \bibinfo {author}
  {\bibfnamefont {T.~P.}\ \bibnamefont {Devereaux}},\ }\href
  {https://doi.org/10.1038/s41467-023-38408-6} {\bibfield  {journal} {\bibinfo
  {journal} {Nat. Commun.}\ }\textbf {\bibinfo {volume} {14}},\ \bibinfo
  {pages} {3129} (\bibinfo {year} {2023})}\BibitemShut {NoStop}%
\bibitem [{\citenamefont {Cullum}\ and\ \citenamefont
  {Willoughby}(1985)}]{CW85}%
  \BibitemOpen
  \bibfield  {author} {\bibinfo {author} {\bibfnamefont {J.~K.}\ \bibnamefont
  {Cullum}}\ and\ \bibinfo {author} {\bibfnamefont {R.~A.}\ \bibnamefont
  {Willoughby}},\ }\href@noop {} {\emph {\bibinfo {title} {Lanczos Algorithms
  for Large Symmetric Eigenvalue Computations}}},\ \bibinfo {series} {Progress
  in scientific computing}, Vol.\ \bibinfo {volume} {I \& II}\ (\bibinfo
  {publisher} {Birkh\"auser},\ \bibinfo {address} {Boston},\ \bibinfo {year}
  {1985})\BibitemShut {NoStop}%
\bibitem [{\citenamefont {Bon\ifmmode~\check{c}\else \v{c}\fi{}a}\ and\
  \citenamefont {Trugman}(2001)}]{BT01}%
  \BibitemOpen
  \bibfield  {author} {\bibinfo {author} {\bibfnamefont {J.}~\bibnamefont
  {Bon\ifmmode~\check{c}\else \v{c}\fi{}a}}\ and\ \bibinfo {author}
  {\bibfnamefont {S.~A.}\ \bibnamefont {Trugman}},\ }\href@noop {} {\bibfield
  {journal} {\bibinfo  {journal} {Phys. Rev. B}\ }\textbf {\bibinfo {volume}
  {64}},\ \bibinfo {pages} {094507} (\bibinfo {year} {2001})}\BibitemShut
  {NoStop}%
\bibitem [{\citenamefont {Lang}\ and\ \citenamefont {Firsov}(1962)}]{LF62}%
  \BibitemOpen
  \bibfield  {author} {\bibinfo {author} {\bibfnamefont {I.~G.}\ \bibnamefont
  {Lang}}\ and\ \bibinfo {author} {\bibfnamefont {Y.~A.}\ \bibnamefont
  {Firsov}},\ }\href@noop {} {\bibfield  {journal} {\bibinfo  {journal} {Zh.
  Eksp. Teor. Fiz.}\ }\textbf {\bibinfo {volume} {43}},\ \bibinfo {pages}
  {1843} (\bibinfo {year} {1962})}\BibitemShut {NoStop}%
\bibitem [{\citenamefont {Chakraborty}\ and\ \citenamefont {Min}(2013)}]{CM13}%
  \BibitemOpen
  \bibfield  {author} {\bibinfo {author} {\bibfnamefont {M.}~\bibnamefont
  {Chakraborty}}\ and\ \bibinfo {author} {\bibfnamefont {B.~I.}\ \bibnamefont
  {Min}},\ }\href@noop {} {\bibfield  {journal} {\bibinfo  {journal} {Phys.
  Rev. B}\ }\textbf {\bibinfo {volume} {88}},\ \bibinfo {pages} {024302}
  (\bibinfo {year} {2013})}\BibitemShut {NoStop}%
\bibitem [{\citenamefont {Wellein}\ and\ \citenamefont {Fehske}(1997)}]{WF97}%
  \BibitemOpen
  \bibfield  {author} {\bibinfo {author} {\bibfnamefont {G.}~\bibnamefont
  {Wellein}}\ and\ \bibinfo {author} {\bibfnamefont {H.}~\bibnamefont
  {Fehske}},\ }\href@noop {} {\bibfield  {journal} {\bibinfo  {journal} {Phys.
  Rev. B}\ }\textbf {\bibinfo {volume} {56}},\ \bibinfo {pages} {4513}
  (\bibinfo {year} {1997})}\BibitemShut {NoStop}%
\bibitem [{\citenamefont {Fehske}\ \emph {et~al.}(1997)\citenamefont {Fehske},
  \citenamefont {Loos},\ and\ \citenamefont {Wellein}}]{FLW97}%
  \BibitemOpen
  \bibfield  {author} {\bibinfo {author} {\bibfnamefont {H.}~\bibnamefont
  {Fehske}}, \bibinfo {author} {\bibfnamefont {J.}~\bibnamefont {Loos}}, \ and\
  \bibinfo {author} {\bibfnamefont {G.}~\bibnamefont {Wellein}},\ }\href@noop
  {} {\bibfield  {journal} {\bibinfo  {journal} {Z. Phys. B}\ }\textbf
  {\bibinfo {volume} {104}},\ \bibinfo {pages} {619} (\bibinfo {year}
  {1997})}\BibitemShut {NoStop}%
\bibitem [{\citenamefont {Chakraborty}\ \emph {et~al.}(2017)\citenamefont
  {Chakraborty}, \citenamefont {Taraphder},\ and\ \citenamefont
  {Berciu}}]{CTB17}%
  \BibitemOpen
  \bibfield  {author} {\bibinfo {author} {\bibfnamefont {M.}~\bibnamefont
  {Chakraborty}}, \bibinfo {author} {\bibfnamefont {A.}~\bibnamefont
  {Taraphder}}, \ and\ \bibinfo {author} {\bibfnamefont {M.}~\bibnamefont
  {Berciu}},\ }\href@noop {} {\bibfield  {journal} {\bibinfo  {journal} {AIP
  Conference Proceedings}\ }\textbf {\bibinfo {volume} {1832}},\ \bibinfo
  {pages} {090025} (\bibinfo {year} {2017})}\BibitemShut {NoStop}%
\bibitem [{\citenamefont {Marchand}\ \emph {et~al.}(2017)\citenamefont
  {Marchand}, \citenamefont {Stamp},\ and\ \citenamefont {Berciu}}]{MSB17}%
  \BibitemOpen
  \bibfield  {author} {\bibinfo {author} {\bibfnamefont {D.~J.~J.}\
  \bibnamefont {Marchand}}, \bibinfo {author} {\bibfnamefont {P.~C.~E.}\
  \bibnamefont {Stamp}}, \ and\ \bibinfo {author} {\bibfnamefont
  {M.}~\bibnamefont {Berciu}},\ }\href {\doibase 10.1103/PhysRevB.95.035117}
  {\bibfield  {journal} {\bibinfo  {journal} {Phys. Rev. B}\ }\textbf {\bibinfo
  {volume} {95}},\ \bibinfo {pages} {035117} (\bibinfo {year}
  {2017})}\BibitemShut {NoStop}%
\bibitem [{\citenamefont {Kohn}(1964)}]{Ko64}%
  \BibitemOpen
  \bibfield  {author} {\bibinfo {author} {\bibfnamefont {W.}~\bibnamefont
  {Kohn}},\ }\href@noop {} {\bibfield  {journal} {\bibinfo  {journal} {Physical
  Review}\ }\textbf {\bibinfo {volume} {133}},\ \bibinfo {pages} {A171}
  (\bibinfo {year} {1964})}\BibitemShut {NoStop}%
\end{thebibliography}
%
\end{document}